%% file: ms.tex
\pdfoutput=1
\documentclass[conference]{IEEEtran}

\input{style.tex}

\begin{document}

\author{
	\IEEEauthorblockN{
		Sanam Ghorbani Lyastani\IEEEauthorrefmark{1},
		Michael Schilling\IEEEauthorrefmark{2},
		Sascha Fahl\IEEEauthorrefmark{3},
		Sven Bugiel\IEEEauthorrefmark{1},
		Michael Backes\IEEEauthorrefmark{4}}
	\IEEEauthorblockA{
		\IEEEauthorrefmark{1}CISPA, Saarland University,
		\IEEEauthorrefmark{2}Saarland University,
		\IEEEauthorrefmark{3}Leibniz University Hannover,
		\IEEEauthorrefmark{4}CISPA Helmholtz Center i.G.
	}
}

\newcommand{\ourtitle}{Studying the Impact of Managers on Password Strength and Reuse}
\title{\ourtitle}

\maketitle

\thispagestyle{plain}
\pagestyle{plain}

\begin{abstract}
Despite their well-known security problems, passwords are still the incumbent authentication method for virtually all online services. To remedy the situation, end-users are very often referred to password managers as a solution to the password reuse and password weakness problems. However, to date the actual impact of password managers on password security and reuse has not been studied systematically.

In this paper, we provide the first large-scale study of the password managers' influence on users' real-life passwords. From 476 participants of an online survey on users' password creation and management strategies, we recruit 170 participants that allowed us to monitor their passwords in-situ through a browser plugin. In contrast to prior work, we collect the passwords' entry methods (e.g., human or password manager) in addition to the passwords and their metrics. Based on our collected data and our survey, we gain a more complete picture of the factors that influence our participants' passwords' strength and reuse. We quantify for the first time that password managers indeed benefit the password strength and uniqueness, however, also our results also suggest that those benefits depend on the users' strategies and that managers without password generators rather aggravate the existing problems.

\end{abstract}

\section{Introduction}

Since several decades textual passwords prevail as the default authentication scheme for virtually all online services~\cite{firstpw}. Despite various research efforts, no ideal alternative scheme has yet been found to replace passwords~\cite{Bonneau:2012:QRP:2310656.2310722,herley2012research} for the simple reasons that, in contrast to their contenders, they are very intuitive to use as well as very inexpensive and effortless to deploy. At the same time, research has again and again demonstrated that passwords perform extremely poor in terms of security~\cite{Morris:1979:PSC:359168.359172}. For instance, various recent attacks exploit the "human factor" of passwords, viz that humans fail to create strong passwords themselves~\cite{6234435,Duermuth2015,197243,passgan,troyhunt_guessing}. Even worse, there is an observable trend towards an increasing number of online services that users register to. This increasing number of required passwords in combination with the limited human capacity to remember passwords lead to the bad practice of re-using passwords across accounts at an alarming rate~\cite{Florencio:2007:LSW:1242572.1242661,naturalhabitat,DBLP:conf/ndss/DasBCBW14,197316}.

In the past, different solutions have been implemented to help users creating stronger passwords, such as password meters and policies, which are also still subject of active research~\cite{Komanduri:2011:PPM:1978942.1979321,Shay:2010:ESP:1837110.1837113,Carnavalet2014FromVW,197243,197177}. Among the most often recommended solutions~\cite{nist800-63b,unichicago,toppwm,192379,185315} to these problems for end-users is technical support in form of password management software. Those password managers come as integrated parts of our browsers, as a plugin to our browsers, or as separate applications. Password managers are being recommended as a solution because they fulfill important usability and security aspects at the same time: They store all the users' passwords so the users do not have to memorize them; they can also help users entering their passwords by automatically filling them into log-in forms; and they can also offer help in creating unique, random passwords. By today, there are several examples for third party password managers that fit this description, such as Lastpass~\cite{lastpass_homepage}, 1Password~\cite{1password_homepage}, and even seemingly unrelated security software, such as anti-virus~\cite{kaspersky_homepage} solutions.

Unfortunately, it has not been sufficiently studied in the past whether password managers fulfill their promise and indeed have a positive influence on password security or not? To break this question down, we are interested in \textit{1)~whether password managers actually store strong passwords that are likely auto-generated by, for instance, password generators, or if they really are just storage where users save their self-made, likely weak passwords?} Further, we are interested whether \textit{2)~users, despite using password managers, still reuse passwords across different websites or if do they use the managers' support to maintain a large set of unique passwords for every distinct service?} Prior works~\cite{197316,naturalhabitat} that studied password reuse and strength in-situ have also considered password managers as factors, but did not find an influence by managers and could not conclusively answer those questions.

\begin{figure*}[t]
    \centering
    \includegraphics[width=.75\linewidth]{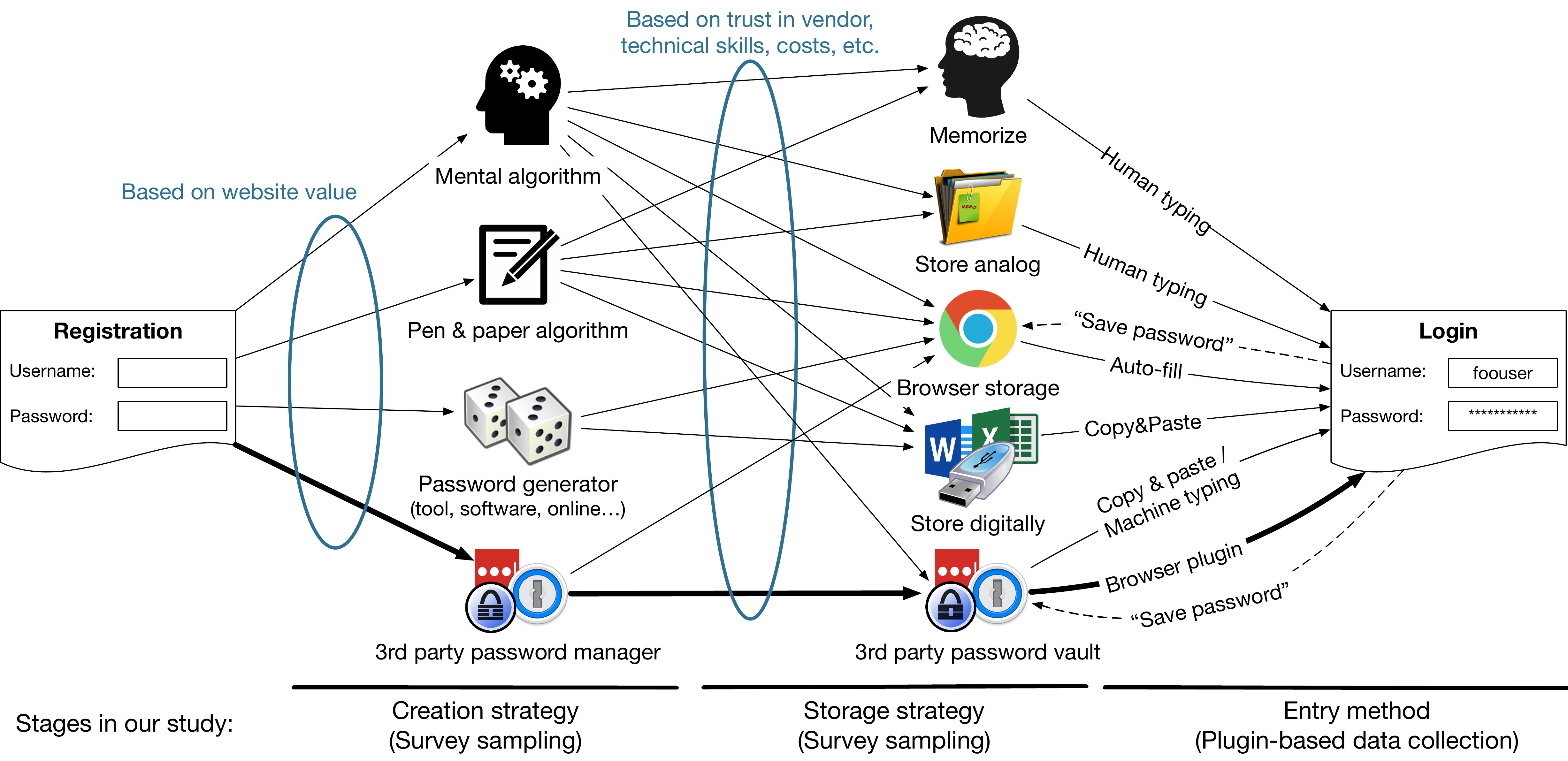}
    \caption{Users' strategies for password creation and storage plus the stages of our study to investigate password managers' influence.}
    \label{fig:methodology_overview}
\end{figure*}

\paragraph*{Our contributions}
We argue that to specifically study the impact of password managers, two important aspects were missing in prior work, which we contribute in this paper. First, previous works considered only the presence of password management software on the user device (potentially even mixed with other factors) and whether a password was auto-filled or not. However, to better distinguish the storage option of a password (i.e., memorized and manually entered, auto-filled by the browser, copy\&pasted, or filled by a browser plugin) a more fine-grained entry method detection is required. Second, users do not axiomatically follow strict workflows for password creation over storage to entry~\cite{Gaw:2006:PMS:1143120.1143127,Hayashi:2011:DSP:1978942.1979326,192379,185315,Tam:2010:PPM:1805058.1805061} (see Figure~\ref{fig:methodology_overview}). For instance, the effort users are willing to invest in creating a new unique and strong password often depends on the privacy-sensitivity of the associated account/website. For creating a new password, the approaches range from mental algorithms (e.g., leetifying a well-known word) over pen\&paper algorithms and separate password generator tools (e.g., websites like \url{https://www.random.org/passwords/} or command line tools like \textit{pwgen}~\cite{pwgen}) to 3rd party password managers (e.g., LastPass, KeePass, 1Password, etc.). Based on different factors, such as technical skills, trust in software vendors, financial expenditure, multi-device support, or others, user resort to different password storage strategies from where the password finds its way via different entry methods into the website form fields. To better study password managers' influence, one has to take the users' creation and storage strategies into consideration as well. In particular, one has to understand if the user pursues primarily a creation strategy based on password manager support and whether there then exists an observable effect of this strategy on the password strength and reuse.

In this paper, we present a study that reflects those considerations (see bottom of Figure~\ref{fig:methodology_overview}). We first recruited 476 participants on Amazon MTurk to conduct a survey sampling to better understand the users strategies for creating and storing passwords, their attitudes towards passwords, and past experiences with password leaks or password managers. From those insights, we identified two distinct groups in our participant pool: users of password managers and users abstaining from technical help in password creation. We were further able to recruit 170 of our participants, 49 of which reported using password managers, for a follow-up study in which our participants allowed us to monitor their passwords through a Google Chrome browser plugin that collected password metrics as well as answers to in-situ questionnaires upon password entry. This gave us detailed information about real-life passwords, including their strength, their reuse, and, for the first time, their entry method (e.g., manually typed, auto-filled, pasted, or entered by a browser plugin) as well as the passwords' context, including user reported value of the password (e.g., loss of social repudiation or financial harm when the password would be leaked).

Based on the combined data from our survey sampling and our plugin-based data collection, we are able to study the factors that influence password strength and reuse from a new perspective. Using exploratory data analysis and statistical testing, including regression models, we show that password managers indeed influence password strength and reuse. In particular, the relation between different entry methods and the password strength depends on the users' entire process of password handling. Using a  workflow that includes technical support from password creation through storage to password entry leads to stronger passwords, while this positive effect cannot be detected when considering the input method individually. For password reuse the picture is even more complex. Passwords entered by a tool that also supports the user during the creation of passwords were significant more unique than passwords entered by hand. But looking at mangers that do not offer this feature, such as Chrome’s auto-fill, we found the opposite effect that those managers even contribute to the problem of password reuse. Lastly, our results also affirm and extend some prior results about reuse as a rampant problem and users’ password behaviors.

\paragraph*{Outline} The remainder of this paper is organized as follows. Section~\ref{sec:relatedwork} gives an overview of related work on studying password security. Section~\ref{sec:methodology} presents the methodology of our study and data collection. We describe the results and analysis of our collected data in Section~\ref{sec:results} and then discuss implications and limitations of study in Section~\ref{sec:discussion}. Finally, we offer some concluding remarks in Section~\ref{sec:conclusion}.

\input{sections/relatedwork.tex}

\input{sections/methodology.tex}

\input{sections/results.tex}

\input{sections/discussion.tex}

\section{Conclusion}
\label{sec:conclusion}

We presented a study on the influence of password managers on password strength and reuse. In our study, we created a more complete view on password managers by combining background data on users, in particular their password storage and creation strategies, with in-situ collected password metrics, most noticeable the passwords' entry methods (e.g., manual entry, browser plugin, copy\&paste). Using  exploratory  data  analysis  and  statistical testing, including regression models, we show that password managers indeed influence password strength and reuse. This is in contrast to prior studies, which did not consider the combination of exact entry method and users' strategies. Whether this influence is beneficial or not, however, depends on how well the manager integrates with the user's creation strategy. We found that users that rely on technical support for password creation had both stronger and more unique passwords, even if entered through other channels than the password manager. We also found, that Google Chrome aggravated the password reuse problem. Future work could explore password managers' influence in other ecosystems, such as Apple's walled-garden ecosystem or mobile password managers. Moreover, the absence of any \textit{ideal} password manager user would warrant further investigation into users' behaviors and strategies and how to further amplify the positive effect of managers (e.g., further decreasing password reuse).

\bibliographystyle{bibstyles/IEEEtranS.bst}
\bibliography{bib} 

\appendices
\input{sections/Appendix.tex}

\end{document}

%% file: style.tex
\usepackage{color}
\usepackage{xcolor}
\usepackage{graphicx}
\usepackage{colortbl}
\usepackage{hhline}

\usepackage{rotating}

\usepackage{pifont}

\usepackage{mdwlist}
\usepackage{enumitem} % customize lists

\usepackage[utf8x]{inputenc}
\usepackage[T1]{fontenc}
\usepackage[english]{babel}

\usepackage{microtype}  % Subliminal refinements towards typographical perfection
\usepackage{xspace}     % Adds a space behind a in-text macro unless the macro is followed by certain punctuation
\usepackage{multicol}   % allows switching between one and multicolumn format on the same page
\usepackage{amssymb}   % symbols for natural numbers
\usepackage{amsfonts}  % symbols for natural numbers
\usepackage{amsmath}   % own operators
\usepackage{stmaryrd}  % The St Mary’s Road symbol font
\usepackage{MnSymbol}  % A Math Symbol Font
\usepackage{pifont}    % http://ctan.org/pkg/pifont

\usepackage{xtab}      % The xtab package enables long tables to be automatically broken at page boundaries
\usepackage{tabularx}  % Tables with a fixed total width and variable column width
\usepackage{multirow}  % table columns spanning multiple rows
\usepackage{booktabs}  % professional tables (advanced rules)

\usepackage{pdfpages}
\usepackage{listings}

\PassOptionsToPackage{hyphens}{url}
\usepackage{url}

\usepackage{pgfplots}
\usepackage{verbatimbox}
\usepackage{cleveref}
\crefname{section}{§}{§§}
\Crefname{section}{§}{§§}

\usepackage{etex}
\usepackage{todonotes}

\usepackage{tikz}
\usetikzlibrary{trees,shapes.geometric,shapes.symbols,chains,calc,positioning,arrows,decorations.pathmorphing,decorations.pathreplacing}

\usepackage{tikz-qtree}
\usetikzlibrary{trees} % this is to allow the fork right path

\definecolor{KWColor}{rgb}{0.37,0.08,0.25}
\definecolor{CommentColor}{rgb}{0.133,0.545,0.133}
\definecolor{StringColor}{rgb}{0,0.126,0.941}

%% file: sections/relatedwork.tex
\section{Related Work}
\label{sec:relatedwork}

Text passwords are since decades~\cite{firstpw} the incumbent authentication scheme for online services~\cite{Hayashi:2011:DSP:1978942.1979326,herley2012research}, and will very likely remain in that position for the foreseeable future. They distinguish themselves from alternative schemes through their very intuitive usage and easy deployment, however, as well as through a pathological inability of users to create strong passwords that withstand guessing attacks~\cite{Bonneau:2012:QRP:2310656.2310722}. Given the permanence of passwords, end-users are commonly referred to technical help in form of password management software~\cite{nist800-63b,unichicago,toppwm,185315} to create strong, unique passwords.

In this paper, we aim to better understand how password managers help users in this task and to try to measure the impact password managers actually have on the current status quo. We do this through a comprehensive study that includes both self-reported user strategies and factors for password creation and storage as well as in-situ collected password metrics and questionnaire answers. To put our approach into the larger context and to provide necessary background information, in this section we give  an overview of prior research on how users select passwords, how users (re-)use passwords, how password strength can be measured, and on dedicated studies of password manager software.

\subsection{Password creation}

Different works have studied the strategies of users and the factors that influence the selection of new passwords. For instance, users create passwords based on something that has relevance to them or has meaning to them~\cite{185315}, and very often passwords are based on a dictionary word~\cite{Inglesant:2010:TCU:1753326.1753384,rinn2015password}.

The effort the user is willing to invest into creating a stronger passwords can depend on different factors. For example, advice and password policies that enforce a certain password composition (i.e., length and character classes) can influence the user~\cite{Yan:2004:PMS:1024867.1025014,Florencio:2007:LSW:1242572.1242661,Inglesant:2010:TCU:1753326.1753384}. Similarly, many websites use password strength meters to provide users real-time feedback on their new password's strength and nudge users into creating stronger passwords~\cite{Egelman:2013:MPG:2470654.2481329,DBLP:conf/uss/UrKKLMMPSVBCC12}. However, often those policies and meters have inconsistent metrics across different websites~\cite{bonneau2010password,wang2015emperor,library978105}, potentially confusing users about what constitutes a strong password~\cite{192379}. Also the value of the password protected account can influence the user. Prior studies~\cite{DBLP:conf/scn/BaileyDP14,Notoatmodjo:2009:PP:1862758.1862770,185315,naturalhabitat} concluded that people try to create strong passwords for accounts that they consider more important, such as banking websites. In particular, users employed password manager for specific matters~\cite{185315} such as just using at a work PC but not at home, or not using password managers for banking websites. Despite their apparent benefits, it is unclear how users \textit{actually} use such password managers and what the exact impact of password managers is on password reuse and password strength.

\subsection{Password strength}

Password strength has been studied for several years. There are different mechanisms that can be use to measure the strength of passwords. The Shannon entropy~\cite{article} equation provides a way to estimate the average minimum number of bits needed to encode a string of symbols, based on the frequency of the symbols. This formula was formerly used by the NIST guidelines~\cite{nist800-63b} to estimate the strength of password based on the length of the passwords. However, more recent research~\cite{Weir:2010:TMP:1866307.1866327,6234435,Dell'Amico:2010:PSE:1833515.1833671,Kelley:2012:GAM:2310656.2310715} argued that \textit{guessability} metrics are a more realistic metric than the common used entropy metrics, and recommendation such as NIST~\cite{nist800-63b} recently picked the results of this line of research up and updated their recommendations accordingly. One of the vital insights from this and other research~\cite{troyhunt_guessing} was that passwords are not chosen randomly but exhibit common patterns and are derived from a limited set of dictionary words.

Measuring a password's guessability has been realized in different ways. Those include Markov models~\cite{conf/ndss/CastellucciaDP12,Duermuth2015}, pattern matching plus word mangling rules~\cite{197177,5207658}, or neural networks~\cite{197243}. Since prior password strength meters were based on the password composition and the resulting entropy, those new approaches also found their way into contending password strength meter implementations~\cite{197177,197243,DBLP:conf/chi/UrAABCCCDNHJM17}. However, varying cracking algorithms or techniques can cause varying password strength results based on configuration, methods, or training data~\cite{Ur:2015:MRA:2831143.2831173}. Also in our study we measure the password strength based on guessability, using the openly available \textit{zxcvbn}~\cite{197177} tool.

\subsection{Password reuse}

Prior work~\cite{185315} have shown that users have an increasing number of online accounts that require creation of a new password. To cope with the task of remembering a large number of passwords, users resort to reusing passwords across different accounts~\cite{DBLP:conf/ndss/DasBCBW14,huntreuse}, creating a situation in which one password leak might affect multiple accounts at once. A large-scale data collection through an instrumented browser~\cite{Florencio:2007:LSW:1242572.1242661} was first to highlight this problem. Since then, newer studies further illustrated the issue of password reuse. For instance, in a combination of measurement study of real leaked passwords and user survey~\cite{DBLP:conf/ndss/DasBCBW14}, 43\% of the participants reused passwords and often a new password was merely a small modification of an existing password. As with password creation, different factors can influence the password reuse. For example, it was shown that the rate of reused passwords increased with the number of accounts~\cite{Gaw:2006:PMS:1143120.1143127}, which is troublesome considering that users accumulate an increasing number of accounts. As with password strength, also the value of the website can affect whether a user creates a unique new password or reuses an existing one~\cite{DBLP:conf/scn/BaileyDP14,naturalhabitat}.

Closest to our methodology are two recent studies~\cite{197316,naturalhabitat} based on data collected from the users' browsers with plugins. Both studies monitored websites for password entries and recorded the password characteristics, such as length and composition, a participant-specific password hash, the web domain (or domain category), as well as meta-information including installed browser plugins, or installed software (e.g., anti-virus software). In case of the newer study~\cite{naturalhabitat}, also hashes of sub-strings of the password were collected as well as a strength estimate using a neural network based password meter~\cite{197243} and whether the password was auto-filled or not. Through this data, both studies had an unprecedented, in-situ insight into user's real password behavior, the factors influencing password reuse, and could show that password reuse, even partial reuse of passwords, is a rampant problem. Further relating to our work, both prior studies also considered the potential influence of password managers, however, could not find any significant effect of password managers on password reuse or strength. However, their studies were not specifically targeted at investigating the impact of password managers, and with our methodology we extend those prior works in two important aspects. First, prior work only considered the presence of password manager and whether auto-fill was used. For our work, we derived a more fine-grained detection of the password entry method, which allows us to distinguish human, plugin-based, auto-fill, or copy\&pasted input to password fields and thus better detection of managed passwords. Second, merely the entry method of password does not reveal its origin (e.g., passwords from a password manager might also be copy\&pasted or saved in the browser's auto-fill). To study the impact of password managers, a broader view is essential that includes the users' password creation strategies in addition to their in-situ behavior.

\subsection{Security and usability of password managers}

Password manager software has also been the subject of research. For instance, human-subject studies~\cite{Karole:2010:CUE:2041036.2041056, Chiasson:2006:USC:1267336.1267337} have shown that password managers might suffer from usability problems and that ordinary users might abstain from using them due to trust issues or because they see a necessity. Like any other software, password manager might also contain vulnerabilities or errors~\cite{184483,zhao2013vulnerability} that can compromise user information, and new guidance for developers of password management software were derived. Also the integration of password manager, in particular the password auto-filling, was scrutinized~\cite{184475,DBLP:conf/ccs/StockJ14} and vulnerabilities discovered that can help an adversary to sniff passwords during the auto-fill process.

%% file: sections/methodology.tex
\section{Methodology}
\label{sec:methodology}

For our study of password managers' impact on password strength and reuse, we use data collected from paid workers of Amazon's crowd-sourcing service \textit{Mechanical Turk}\footnote{\url{https://www.mturk.com/}}. We collected the data in three different stages: 1)~an initial survey sampling, 2)~collection of \textit{in-situ} password metrics, and 3)~an exit survey. In the following, we describe those three stages in more detail.

\paragraph*{Ethical concerns}

The protocols implemented in those two stages were approved by the ethical review board\footnote{\url{https://erb.cs.uni-saarland.de/}} of the faculty of Mathematics and Computer Science at Saarland University. We also took the strict German data and privacy protection laws into account for collecting, processing, and storing any participant information. Further, we followed the guidelines for academic requesters outlined by MTurk workers~\cite{requesterguidelines}.
All server-side software (i.e., a LimeSurvey Community Edition software and a self-written server application for our plugin-based data collection) was self-hosted on a maintained and hardened university server. Web access to the server was secured with an SSL certificate issued by the university's computing center and all further access was restricted to the department's intranet and only made available to maintainers and collaborating researchers. Participants could leave the study at any time during the two stages.

\subsection{Password survey}
\label{sec:samplingsurvey}

In our survey sampling, we asked the participants about their general privacy attitude, their attitude towards passwords, their skills and strategies for creating and managing passwords, as well as basic demographic questions. Those information enable us, on the one hand, to gain a general overview of common password creation and storage strategies \textit{in the wild}. On the other hand, those information help us in detecting and avoiding any potential biases in the later stages of our study. The full survey contained 31--34 questions, depending on conditional questions, categorized in 6 different groups (see Appendix~\ref{sec:questionnaire}).

We first asked for their privacy attitude using the standard Westin index~\cite{Kumaraguru_Cranor:2005}. However, since the Westin index has been shown to be an unreliable measure of the actual privacy-related actions of users~\cite{selldna}, we also asked about the participants' attitude towards passwords (e.g., whether they consider passwords to be futile in protecting their privacy). This should help in better understanding if participants are actually motivated to put effort in creating stronger and unique passwords. We further asked about the participants' strategies for password creation and management in order to get a more complete picture about the possible origins of passwords in our dataset. Lastly, we collected demographic information about the participants.

All qualitative answers (e.g., free text answers to \textit{Q9} or \textit{Q22} in Appendix~\ref{sec:questionnaire}) were independently coded in a bottom-up fashion by two researchers. For the coding tasks, the researcher achieved an initial agreement between 95.6\% (\textit{Q9}) and 97.1\% (\textit{Q22}) and all differences could be resolved in agreement.

Participation in the survey was open to any MTurk worker that fulfilled the following criteria, which we copied from MTurk-based studies in psychological research: the worker has to be located in the US and the number of previously approved tasks has to be at least 100 or at least 70\% all of tasks. By using MTurk, we ensured that all participants were at least 18 years old. The estimated time for answering the survey was 10--15 minutes and we paid workers \$4 for participation.

In total, 505 MTurk workers participated in our survey between August 2017 and October 2017. After discarding responses that failed attention test questions~\cite{Huang2015}, were answered too fast to be done thoughtfully, or that were duplicates (e.g., same human worker with different IDs), we ended up with 476 valid responses.

Lastly, we also asked whether the participant would be willing to participate in a follow-up study, in which we measure in an anonymized, privacy-protecting fashion the strength and reuse of their passwords. Only participants that indicated interest in the follow-up study were considered potential candidates for our Chrome plugin-based data collection. Only 21 workers were not interested.

\subsection{Chrome plugin based data collection}
\label{sec:chromeplugin}

To collect in-situ data about passwords, including their strength, reuse, entry method, and domain, we created a Chrome browser plugin that monitors the input to password fields of loaded websites and then sends all collected metrics back to our server once the user logs into the loaded website. We distributed our Chrome plugin via the Google Web Store to invited participants. The plugin was unlisted in the Google Web Store, so that only participants to which we sent the link to the plugin store website were able to install it. Our primary selection criterion for participant selection was that they use Chrome as their primary browser and are not using exclusively mobile devices (smartphones and tablets) to browse the web; besides that we aimed for an unbiased sampling from the participants pool with respect to the participants' privacy attitude, attitude towards passwords, demographics, and usage of password managers. Between September and October 2017, we invited 364 participants from the survey sampling via MTurk to the study, of which 174 started and 170 finished participation. Participants that finished the task were compensated with \$20.

To monitor password entries, we follow and extend prior approaches on password monitoring~\cite{naturalhabitat, 197316,Florencio:2007:LSW:1242572.1242661}. Our plugin searches in the DOM tree of the currently loaded website for \textit{input} elements with type \textit{'password'} and registers different event listeners for those identified elements, which monitor key presses, key down/up events, or paste events. To detect login attempts that would submit the entered password to the website, the plugin registers listeners for different forms of webform submission (e.g., inputs with type \textit{button} or \textit{submit}, or \textit{div} elements with special roles, such as \textit{button}) as well as for pressing the \textit{enter} key in the detected password field. A limitation of this approach is that the website structure has to follow common programming patterns (e.g., input types and submission forms). During our pilot testing with Alexa's top 100 websites, we observed only three websites that failed this assumption: two adult entertainment websites and msn.com. Once a login attempt is detected, the plugin analyzes the recorded events and input to the password field. As a result of that analysis, the plugin collects the following metrics about the entered passwords:

\subsubsection*{Composition} The length of the password as well as the frequency of each character class.

\subsubsection*{Strength} The password strength measured in Shannon and NIST entropy as well as zxcvbn score. Shannon and NIST entropy have been used in prior works~\cite{Fahl:2013:EVP:2501604.2501617,197316,Egelman:2013:MPG:2470654.2481329} as measure of password strength and complexity and are collected primarily to be backwards compatible in our analysis with prior research. However, since entropy has been shown to be a poor measurement of the actual "crackability" of the password~\cite{Weir:2010:TMP:1866307.1866327}, we use the zxcvbn~\cite{197177} score as the more realistic estimator of the password strength in our analysis.\footnote{Unfortunately, the fully trained neural network based strength estimator of \cite{naturalhabitat,197243} was not publicly available.} Zxcvbn\footnote{\url{https://github.com/dropbox/zxcvbn}} by Dropbox~\cite{197177} is password strength estimator that uses pattern matching (e.g., repeats, sequences, keyboard patterns), (common) password dictionaries (including leaked passwords, names, English dictionary words), and mangling rules (e.g., l33t speak) to estimate the crackability of passwords. It estimates every password's strength on a scale from 0 (weakest) to 4 (strongest). In our plugin we used the zxcvbn library~\cite{github_zxcvbn} with its default settings. For instance, the password \lstinline{!@#$%^&*()} is estimated with score 1, since it consists of a straight row of keys, and \lstinline{AiWuutaiveep9}, which was randomly generated, with score 4.

 \begin{table}[t]
     \centering
     \caption{Zxcvbn scores for 200~million unique passwords from hashes.org. Guesses are in log10.}
     \label{tab:zxcvbn_scoring}
     \scriptsize
     \begin{tabular}{l|r|r|r|r|r|r|r|r}
          \multicolumn{1}{c}{} & \multicolumn{1}{c|}{} & \multicolumn{7}{c}{\cellcolor[gray]{.9}Guesses (in log10)}\\
          Score & Count       & Mean  & SD    & Min   & 25\%   & 50\%  & 75\%  & Max \\\hline
          0     & 122,296     & 2.69  & 0.42  & 0.30  & 2.48   & 2.92  & 3.00  & 3.00 \\
          1     & 34,496,960  & 5.34  & 0.59  & 3.00  & 5.00   & 5.44  & 5.87  & 6.00 \\
          2     & 69,090,776  & 7.15  & 0.66  & 6.00  & 6.61   & 7.00  & 7.87  & 8.00 \\
          3     & 57,256,840  & 8.87  & 0.65  & 8.00  & 8.28   & 8.87  & 9.36  & 10.00 \\
          4     & 39,789,207  & 12.51 & 2.29  & 10.00 & 11.00  & 12.00 & 13.36 & 32.00 \\
     \end{tabular}
 \end{table}

To better understand zxcvbn's scoring, we used zxcvbn to score 200~million unique passwords collected from \url{hashes.org}, where we measured the zxcvbn score and the corresponding guesses in log10. The results in Table~\ref{tab:zxcvbn_scoring} show that each zxcvbn score has a corresponding bin of guesses, e.g., passwords with score 2 require between $10^{3}$--$10^{6}$ guesses, and passwords that need more than $10^{10}$ guesses are rated with score 4.

\subsubsection*{Website category} The category of the website domain according to the \textit{Alexa Web Information Service}~\cite{awsi_devguide}.

\subsubsection*{Entry method} The method through which the password was entered, such as \textit{human}, \textit{Chrome built-in password manager}, \textit{copy\&paste}, \textit{3rd party password manager plugin}, or \textit{external password manager program}. The heuristics to detect the entry method are described separately in Section~\ref{sec:entrymethod}.

\subsubsection*{In-situ questionnaire} The participant's answers to a short questionnaire about the entered password and website (see Section~\ref{sec:instructions}). In particular, we ask the participants about the website's \textit{value} for their privacy. Other studies used the website category as a proxy for this value~\cite{naturalhabitat} and in our study we wanted first-hand knowledge.

\subsubsection*{Hashes} Adapting the methodology of \cite{naturalhabitat,197316}, we collect the hash of the entered password as well as the hash of every 4-character sub-string of the password. We use a keyed hash (i.e., PBKDF2 with SHA-256), where the key is generated and stored at the client side and never revealed to us, and only collected the \textit{first half} of the resulting hash. This still allows identification of (partially) reused passwords \textit{per participant} with a negligible error chance and makes a trade-off in favor of the participants' privacy (see also Section~\ref{sec:privacyconcerns}). In the remainder of this paper, we will use the notions of \textit{partial}, \textit{exact}, and \textit{partial-and-exact} reuse introduced in \cite{naturalhabitat}. Exactly reused passwords are identical with another password, partially reused passwords share a substring with another password, and partially-and-exactly reused passwords have both of those characteristics. Like \cite{naturalhabitat,197316}, we cannot compare passwords across participants.

\subsubsection{Detecting the entry method}
\label{sec:entrymethod}

\begin{figure}[t]
    \centering
    \includegraphics[width=\linewidth]{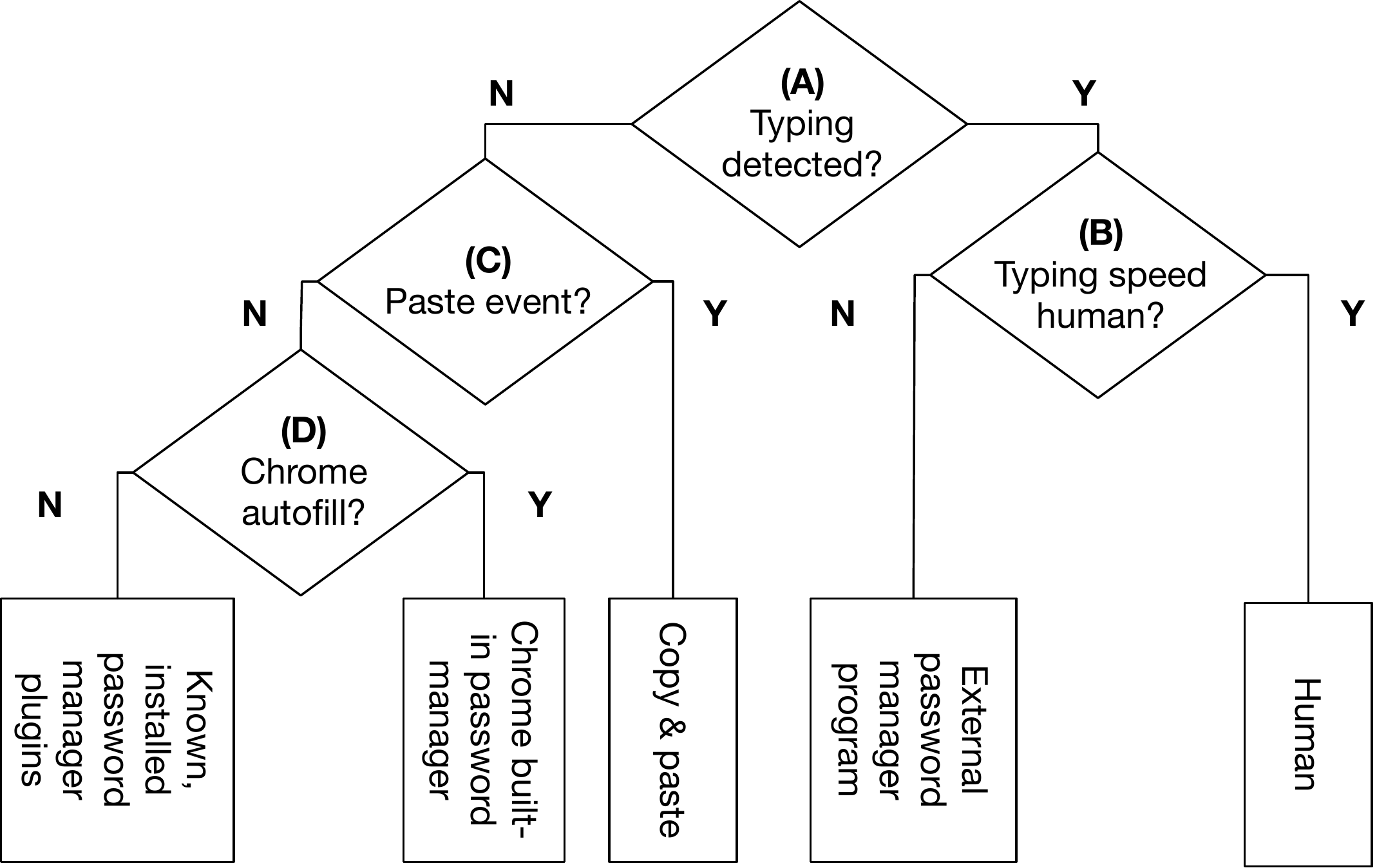}
    \caption{Decision tree of our plugin to detect password entry methods}
    \label{fig:entrymethod}
\end{figure}

Detection of the password entry method uses the different listeners registered by our plugin and follows the decision tree depicted in Figure~\ref{fig:entrymethod}. If our plugin detects any kind of typing inside the password field (\textbf{(A)=Y} in Figure~\ref{fig:entrymethod}) and the typing speed is too fast to be from a human typist (\textbf{(B)=N}), we conclude that an external password manager program (such as KeePass) mimics a human typist by "replaying" the keyboard inputs of the password. Otherwise (\textbf{(B)=Y}), we assume a manually entered password. As threshold between human and an external program mimicking a human typist, we set an average key press time of 30~ms. This is based on the observation that external programs usually do not consider mimicking the key press time (i.e., duration of keypress), while some of them enter the password character-wise with varying speeds. In case there was no typing in the password field detected (\textbf{(A)=N}) and a paste event was observed (\textbf{(C)=Y}), we consider the password to be pasted into the field, either by a human or by an external program. In either of those two cases the password is managed external to the browser in digital form. If no paste event was detected (\textbf{(C)=N}) and the Chrome auto-fill event was observed, this indicates that Chrome filled the password field from its internal, built-in password manager (i.e., \textit{"save password feature"}). If Chrome auto-fill has not filled the password field (\textbf{(D)=N}), we check the list of installed plugins for known plugins of password manager solutions, such as LastPass, 1Password, etc. Chrome plugins are identified through a 32 characters long ID that can be retrieved from Google's Chrome Web Store (e.g., \lstinline{hdokiejnpimakedhajhdlcegeplioahd} for the LastPass plugin). Our plugin checks for eight well-known password manager plugins (see Table~\ref{tab:known_plugins}) and reports the ones installed in the participant's browser, or an "unknown" value in case none of those eight was found.

 \begin{table}[t]
     \caption{UUIDs of known plugins that are detected by our study plugin}
     \label{tab:known_plugins}
     \centering
     \begin{tabular}{l|r} \toprule
         \textbf{Name} & \textbf{UUID} \\\midrule
         Dashlane & \texttt{fdjamakpfbbddfjaooikfcpapjohcfmg}\\
         LastPass & \texttt{hdokiejnpimakedhajhdlcegeplioahd}\\
         1Password & \texttt{aomjjhallfgjeglblehebfpbcfeobpgk}\\
         Roboform & \texttt{pnlccmojcmeohlpggmfnbbiapkmbliob}\\
         Enpass & \texttt{kmcfomidfpdkfieipokbalgegidffkal}\\
         Zoho Vault & \texttt{igkpcodhieompeloncfnbekccinhapdb}\\
         Norton Identity Safe & \texttt{iikflkcanblccfahdhdonehdalibjnif}\\
         KeePass & \texttt{ompiailgknfdndiefoaoiligalphfdae}\\
         \bottomrule
     \end{tabular}
 \end{table}

In this decision tree, we make the assumption that the user does not enter the password with a mixture of the different entry methods (e.g., pasting a part of the password and complementing it with typing). For instance, we do not check if there was a paste event when typing was detected. Such mixture of entry methods will hence result in misclassification of the detected entry method. However, we assume that such user behavior is too rare to affect our results statistical significantly.

\subsubsection{Participant instructions}
\label{sec:instructions}

\begin{figure}[t]
    \centering
    \includegraphics[width=\linewidth]{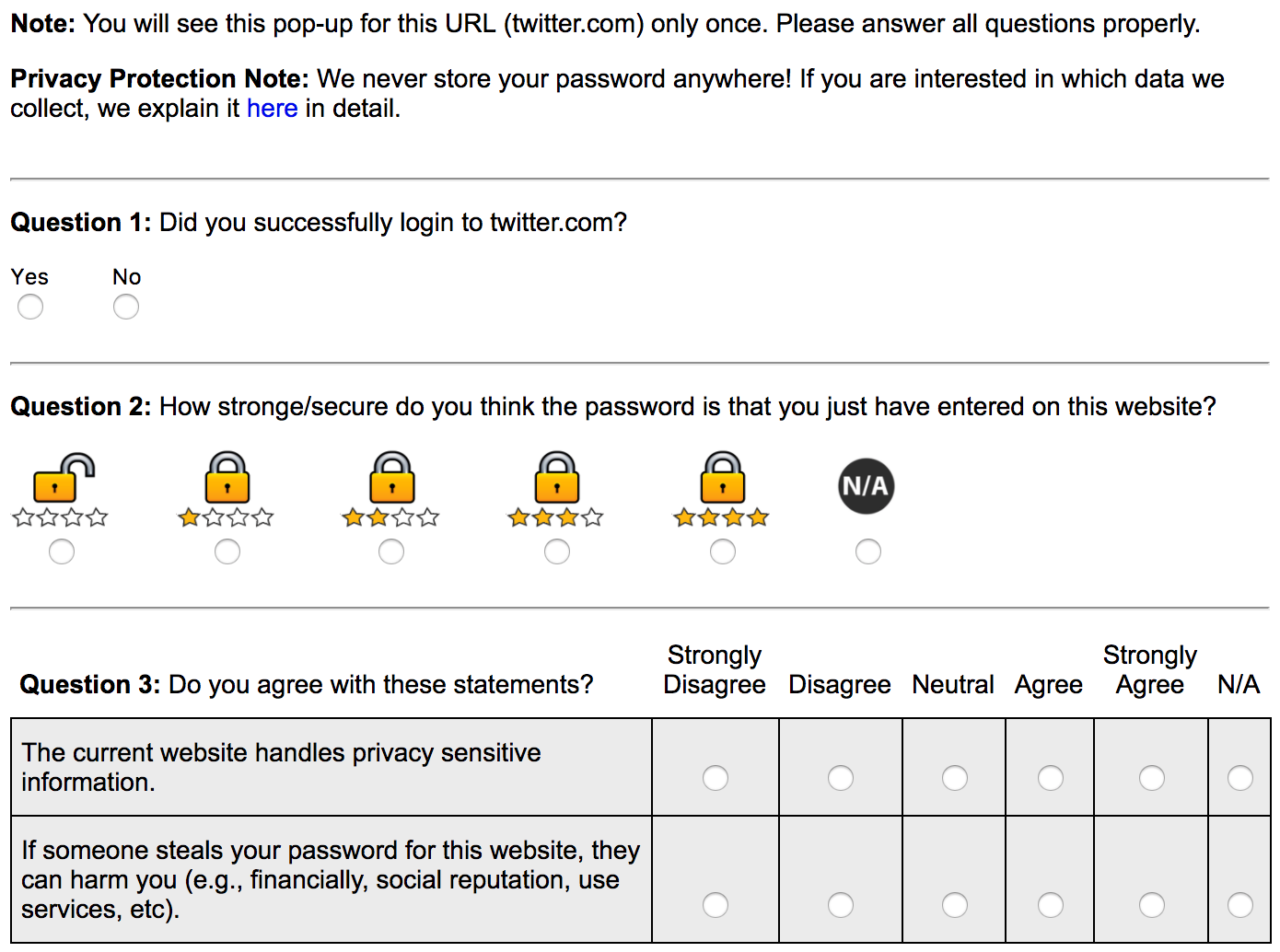}
    \caption{In-situ questionnaire when a login to new website is detected.}
    \label{fig:popup1}
\end{figure}

We provided our participants with a project website that gave a step-by-step introduction on how install our plugin, how to set it up, how to use it, and how to remove it post-participation. Google's Web Store provided our participants with a very comfortable way of adding the plugin to their browser. To set the plugin up, participants had to simply enter their MTurk worker ID into the plugin. The worker ID was used as a pseudonym throughout this study to identify data of them same participant. After setup, the plugin starts monitoring the users' websites for password entries. For every newly detected domain to which a password was submitted, our plugin asked the participant to answer a short three question questionnaire about the participants' estimate of the website's value, the participants' strength estimate of the just entered password, and whether the login was successful (see Figure~\ref{fig:popup1}). Every participant was instructed to use the plugin for four days, after which the plugin released a completion code to be entered into the task on MTurk to finish participation and collect the payment. We also instructed participants to act naturally and not change their usual behavior during those four days in order to maximize the ecological validity of our study. The only exceptions from the usual behavior were the installation of our plugin and a request to re-login to all websites where they have an account in order to ensure a sufficient enough quantity of collected data.

\subsubsection{Addressing privacy concerns}
\label{sec:privacyconcerns}

A particular consideration of our study design were the potential privacy concerns of our participants. Since we essentially ask our participants to install a key-logger that monitors some of the most privacy-sensitive data, this might make participants wary of the study and repel them from participating. Due to the lack of any form of in person interview or consultation between the researchers and the participants, we tried to address those concerns through a high level of transparency, support, and collecting only the minimal amount of data in a privacy-protecting fashion, which also follows the guidelines for academic requesters~\cite{requesterguidelines} as outlined by the MTurk community.

First, we explained on our project website exactly the motivation behind our study and why acting naturally is important for our results. In this context, we provided a complete list of all data that our plugin collects, for which purpose, and why this data collection does not enable use to steal (or break) the participants passwords. We also answered all participants' questions in this regard that were sent to us via email or posted in known MTurk review and discussion forums. We received feedback from workers that this level of openness has convinced them to participate in the study.

Second, to further ensure transparency, we asked concerned workers with IT background to review the code of our plugin, which is distributed in an authenticated way via the Google Web Store. However, to the best of our knowledge, no worker that investigated the code has publicly reported about it.

Third, we limited the extent of the collected data to the necessary minimum while still being able to study password managers' impact. For instance, we only collect the first successful login to any website, thus abstaining from monitoring the participants' browsing behavior. Further, we only collect the website category and not the website URL. This trade-off between participants' privacy and data accuracy comes at the cost of data completeness, since the category of any URL not in the category database of our plugin is unknown to us. Our plugin currently contains the category for the top 28,651 web domains in Alexa at the time the study was conducted.\footnote{This is the number of web domains in the top 100K list, for which a category was assigned by Alexa.} Lastly, similar to \cite{naturalhabitat}, we collect the password hashes such that we are unable to recover the original password, while still being able to detect (partially) reused passwords for each participant.

\begin{figure}[t]
    \centering
    \includegraphics[width=\linewidth]{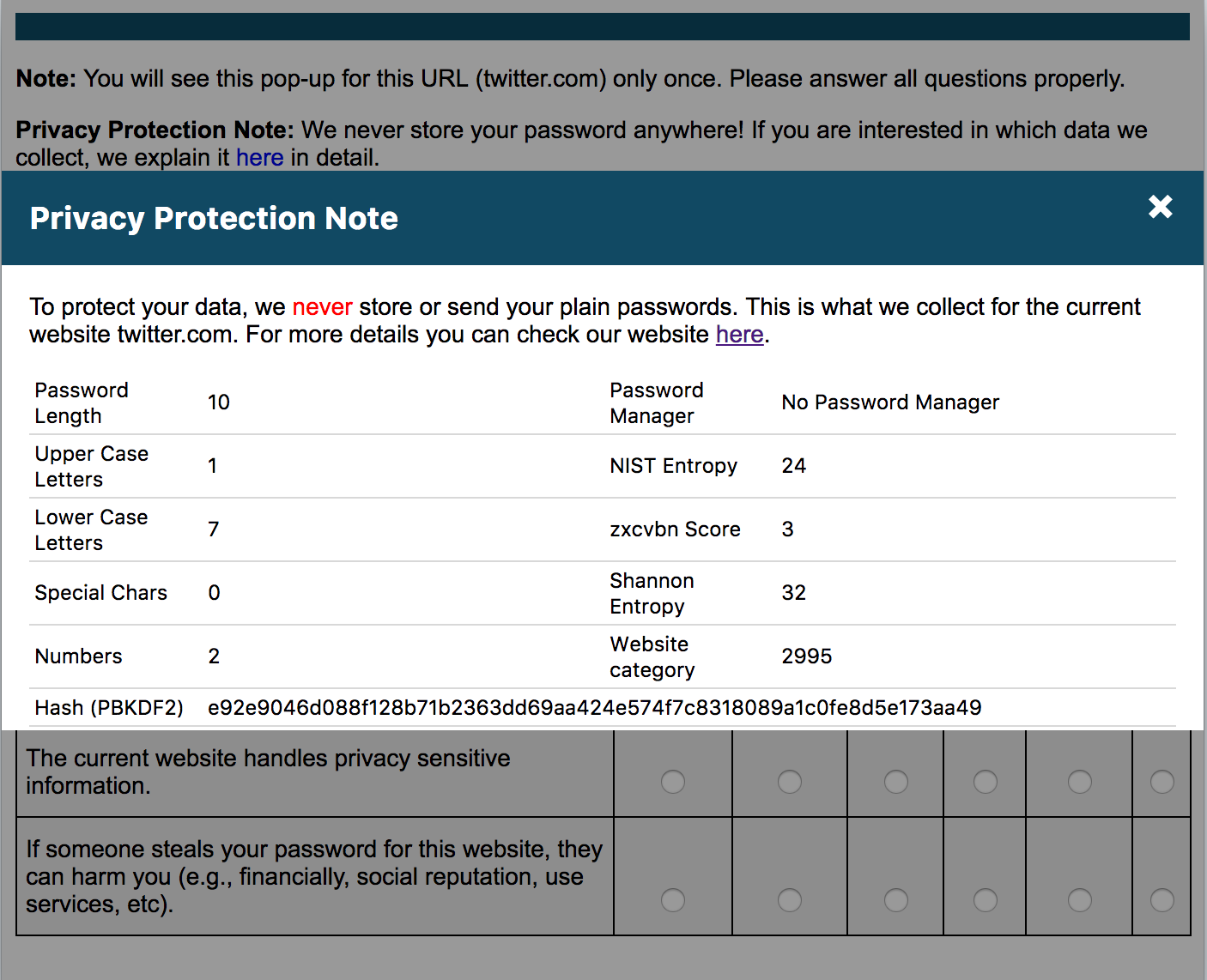}
    \caption{Notice by our plugin about the collected data for the current website.}
    \label{fig:popup2}
\end{figure}

Fourth, every participant could inspect the collected data prior to sending them to our server. Figure~\ref{fig:popup2} shows the information screen for participants, which lists the collected information. Moreover, the data collection can be avoided for highly sensitive websites the participant does not want to be collected by simply not answering the questionnaire. Only data for websites where the questionnaire (in Figure~\ref{fig:popup1}) was answered by the participant were collected by the plugin.

\subsection{Exit survey}

In the last stage of our study, we conducted a short, four question exit survey to better understand the reasons why users do not adopt external password manager software. In particular, we asked for the concrete reason why they do not use such a software, whether they have in the past, and if so, why they stopped using it. For this survey, we invited participants that noted in their survey that they do not use any extra password manager software and for which we could corroborate this claim in our collected data. We invited them via email through MTurk. From 113 invited workers, 109 answered the exit survey. Each worker was compensated with \$1.50 for participation.

%% file: sections/results.tex
\section{Studying Password Managers' Impact}
\label{sec:results}

In this section, we analyze our collected data, but leave the discussion of the results for Section~\ref{sec:results}. We start with our participants' demographics. We then present a brief overview of our participants' password reuse and strength in general and afterwards introduce a grouping of our participants based on their creation strategy. We then present a short case-study of LastPass users. Finally, we study the impact of different management and creation strategies on the password reuse and strength by exploring correlations between those factors.

\subsection{Demographics}

Table~\ref{tab:demographics} provides an overview of the demographics of our participants that answered our survey on passwords, that we invited to the plugin-based study, and that participated in the plugin-based data collection. Noticeable is that we invited participants in equal parts from every demographic group and that also every demographic group participated in almost equal parts in the plugin-based data collection. We us a Mann-Whitney rank test~\cite{discoverstatistics} to test for significant differences between the demographic distributions of the 476 participants in the survey sampling and the 170 participants in the plugin-based study, and could not find any statistically significant ($p<.05$) differences between those two groups. In general, our participants' demographics are closer to the commonly observed demographics of qualitative studies in university settings than to the demographics of the 2010 US census~\cite{uscensus}. Our participants' number is skewed towards male participants (57.6\% identified themselves as male). Also, our participants covered an age range from 18 to more than 70 years, where our sample skews to younger participants (75.2\% of our study participants are younger than 40) as can be commonly observed in behavioral research, including password studies and usable security. The majority of our participants had no computer science background (80.88\%) and was English speaking (98.3\%). Most of the participants identified themselves as of white/Caucasian ethnicity (74.6\%). The participants also covered a range of educational levels, where 14.3\% are high school graduates, 16.6\% having an associate's degree, 36.6\% having completed a Bachelor's degree, 0.4\% having a doctoral degree, and 6.7\% having completed a graduate or professional degree. Further, 80.9\% of our participants reported using Chrome as their primary browser (see Table~\ref{tab:browsershare}).

\begin{table}[t]
    \caption{Demographics of our survey sampling, of our selection of invited participants for the follow-up study, and of participants that finished the plugin-based study. Percentages indicate the fraction w.r.t.~initial size in the survey sampling.}
    \label{tab:demographics}
    \centering
    \scriptsize
    \begin{tabular}{l|r|r|r}\toprule
    & Survey & Invited to study & Participated \\
    \hline
    Number of participants & 476 & 364 & 170 \\\hline
    \rowcolor[gray]{.9}
    \multicolumn{4}{l}{Gender}\\\hline
    Female      & 200   & 156 (78.0\%)  & 73 (36.5\%) \\
    Male        & 274   & 208 (75.9\%)  & 97 (35.4\%) \\
    Other       & 1     & 0             & 0 \\
    No answer   & 1     & 0             & 0 \\
    \hline
    \rowcolor[gray]{.9}
    \multicolumn{4}{l}{Age group}\\\hline
    18--30      & 180   & 139 (77.2\%)  & 64 (35.6\%) \\
    31--40      & 178   & 135 (75.8\%)  & 63 (35.4\%) \\
    41--50      & 71    & 58 (81.7\%)   & 32 (45.1\%) \\
    51--60      & 35    & 24 (68.6\%)   & 8 (22.9\%) \\
    61--70      & 11     & 7 (63.6\%)   & 2 (18.2\%) \\
    $\geq$71    & 1     & 1 (100\%)     & 1 (100\%)\\
    \hline
    \rowcolor[gray]{.9}
    \multicolumn{4}{l}{Computer science background}\\\hline
    Yes         & 91    & 64 (70.3\%)   & 27 (29.7\%) \\
    No          & 385   & 300 (77.9\%)  & 143 (37.1\%) \\
    \hline
    \rowcolor[gray]{.9}
    \multicolumn{4}{l}{Native language}\\\hline
    English     & 468   & 358 (76.5\%)  & 167 (35.7\%) \\
    Other       & 8     & 6 (75.0\%)    & 3 (37.5\%) \\
    \hline
    \rowcolor[gray]{.9}
    \multicolumn{4}{l}{Education level}\\\hline
    Less than high school   & 3     & 3 (100\%)     & 1 (33.3\%) \\
    High school graduate    & 68    & 53 (77.9\%)   & 26 (38.2\%) \\
    Some college, no degree & 117   & 85 (72.6\%)   & 34 (29.1\%) \\
    Associate's degree      & 79    & 64 (81.0\%)   & 34 (43.0\%) \\
    Bachelor degree         & 174   & 133 (76.4\%)  & 62 (35.6\%) \\
    Ph.D                    & 2     & 1  (50.0\%)   & 1  (50.0\%) \\
    Graduate/prof. degree   & 32    & 25 (78.1\%)   & 12 (37.5\%) \\
    Other                   & 1     & 0             & 0  \\ 
    \hline
    \rowcolor[gray]{.9}
    \multicolumn{4}{l}{Ethnicity}\\\hline
    White/Caucasian               & 355   & 274 (77.2\%)   & 123 (34.6\%) \\
    Black/African American        & 50    & 38 (76.0\%)    & 25 (50.0\%) \\
    Asian                         & 31    & 23 (74.2\%)    & 9 (29.0\%) \\
    Hispanic/Latino               & 27    & 21 (77.8\%)    & 12 (44.4\%) \\
    Native American/Alaska        & 1     & 0              & 0  \\
    Multiracial                   & 7     & 5 (71.4\%)     & 1 (14.3\%) \\
    Other                         & 5     & 3 (60.0\%)     & 0  \\
    \bottomrule
   \end{tabular}
\end{table}

\begin{table}[t]
    \caption{Primary browsers used, as reported by 476 our participants. Percentages in relation to overall number of participants.}
    \label{tab:browsershare}
    \centering
    \scriptsize
    \begin{tabular}{l|c|c|c|c|c|c}
    \textbf{Browser} & Chrome & Firefox & Safari & Opera & IE/Edge & Other \\\midrule
    \textbf{Share}   & 385 & 71  & 7  & 6  & 1  & 6 \\
    & (80.9\%) & (14.9\%) & (1.5\%) & (1.3\%) & (0.2\%) & (1.3\%)
    \end{tabular}
\end{table}

Since our study effectively asks participants to install a password-logger, we were concerned with a potential opt-in bias towards people that have low privacy concerns or consider passwords as ineffective security measures. To this end, we included the three questions of Westin's Privacy Segmentation Index~\cite{Kumaraguru_Cranor:2005} (\textit{Q1} in Appendix~\ref{sec:questionnaire}) to capture our participants' general privacy attitudes (i.e., fundamentalists, pragmatists, unconcerned). We further added two questions specifically about our participant's attitude about passwords (see \textit{Q4} in Appendix~\ref{sec:questionnaire}), e.g., if passwords are considered a futile protection mechanism or important for privacy protection. Table~\ref{tab:attitude} summarizes the results of those questions. Only a minority of 86 of our survey participants are privacy unconcerned and the majority of 365 participants believe in the importance of passwords as a security measure. Almost a third of our survey participants experienced a password leak in the past. For our study we sampled in almost equal parts from those different groups. Using a Mann-Whitney rank test, we could not find any statistically significant differences between the survey and study participants' distribution of privacy and password attitudes/experiences. Thus, we argue that the risk of an opt-in bias towards either end of the spectrum for privacy and password attitude is unlikely.

\begin{table}[t]
    \caption{Privacy attitude (according to Westin index), attitude about passwords, and prior experience with password leakage among our participants in the survey, list of selected participants for the follow-up study, and of participants that actually participated in the data collection.}
    \label{tab:attitude}
    \centering
    \begin{tabular}{l|r|r|r}\toprule
    & Survey & Invited to study & Participated \\
    \hline
    \rowcolor[gray]{.9}
    \multicolumn{4}{l}{Privacy concern (Westin index)}\\\hline
    Fanatic     & 217   & 167 (77.0\%)  & 66 (30.4\%) \\
    Unconcerned & 86    & 56 (65.1\%)   & 31 (36.0\%) \\
    Pragmatist  & 173   & 141 (81.5\%)  & 73 (42.2\%) \\
    \hline
    \rowcolor[gray]{.9}
    \multicolumn{4}{l}{Attitude about passwords}\\\hline
    Pessimist   & 9     & 8 (88.9\%)    & 3 (33.3\%) \\
    Optimist    & 365   & 279 (76.4\%)  & 132 (36.2\%) \\
    Conflicted  & 102   & 77 (75.5\%)   & 35 (34.3\%) \\
    \hline
    \rowcolor[gray]{.9}
    \multicolumn{4}{l}{Prior password leak experienced}\\\hline
    No              & 190   & 151 (79.5\%)  & 72 (37.9\%) \\
    Yes             & 148   & 111 (75.0\%)  & 58 (39.2\%) \\
    Not aware of    & 138   & 102 (73.9\%)  & 40 (29.0\%) \\
    \bottomrule
    \end{tabular}
\end{table}

\subsection{General password statistics}

Tables~\ref{tab:pwsummary} and \ref{tab:em_breakdown} provide summary statistics of all passwords collected by our plugin. We collected from our 170 participants 1,045 unique passwords and 1,767 password entries in total. That means, that our average participant entered passwords to 10.40 distinct domains with a standard deviation of 5.52 and median of 9. The lowest number of domains per participant is 1 and the highest number is 27, where the first quartile is 6 and the third quartile is 14. Those numbers are hence slightly lower than those reported in recent studies~\cite{naturalhabitat}. When considering only unique passwords, our average participant has only 6.15 passwords, indicating that passwords are reused frequently~\cite{Gaw:2006:PMS:1143120.1143127,naturalhabitat}. Our participants entered their passwords on average with 2.24 different entry methods, where 24 participants used only one  method and 8 participants used four different methods. Including all passwords, our participants reused on average 70.56\% of their passwords, where exact-and-partial reuse is most common with 36.46\% of all passwords. Interestingly the minimum and maximum in all reuse categories is 0\% and 100\%, respectively, meaning that we have participants that did not reuse any of their passwords as well as participants that exactly, partially, or partially-and-exactly reused all of their passwords. The average password in our dataset had a length of 9.61 and was composed of 2.52 character classes. The average zxcvbn score was 2.20, where the participant with the weakest passwords had an average of 0.67 and the participant with the strongest passwords an average of 4.00. Like prior work~\cite{197316}, we observe a significant correlation between password strength and reuse (chi-square test: $\chi^2=75.48$, $p<.001$).

\begin{table}[t]
    \caption{Summary statistics for all 170 participants in our plugin-based data collection. Like~\cite{naturalhabitat}, we first computed means for each participant and then computed the mean, median, standard deviation, and min/max values of those means.}
    \label{tab:pwsummary}
    \centering
    \scriptsize
    \begin{tabular}{l|r|r|r|r|r}
    \toprule
\textbf{Statistic} & \textbf{Mean} & \textbf{Median} & \textbf{SD} & \textbf{Min} & \textbf{Max}\\\midrule
Number of passwords & 10.39 & 9.00 & 5.52 & 1.00 & 27.00\\
Entry methods & 2.24 & 2.00 & 0.75 & 1.00 & 4.00 \\\hline
\rowcolor[gray]{.9}\multicolumn{6}{l}{Percentage reused passwords}\\\hline
Non-reused & 29.44\% & 21.58\% & 28.25\% & 0.00\% & 100.00\% \\
Only-exact-reused & 15.72\% & 0.00\% & 24.43\% & 0.00\% & 100.00\% \\
Only-partially-reused & 18.38\% & 11.11\% & 19.88\% & 0.00\% & 100.00\% \\
Exact-and-partial reused & 36.46\% & 38.75\% & 30.88\% & 0.00\% & 100.00\% \\\hline
\rowcolor[gray]{.9}\multicolumn{6}{l}{Password composition}\\\hline
Length & 9.61 & 9.29 & 1.72 & 6.33 & 16.86  \\
Character classes & 2.52 & 2.50 & 0.58 & 1.00 & 3.94 \\
Digits & 2.54 & 2.38 & 1.24 & 0.25 & 6.73  \\
Uppercase letters & 0.85 & 0.67 & 0.81 & 0.00 & 4.62 \\
Lowercase letters & 5.92 & 5.72 & 1.96 & 1.67 & 15.50 \\
Special characters & 0.30 & 0.10 & 0.54 & 0.00 & 5.19 \\\hline
\rowcolor[gray]{.9}\multicolumn{6}{l}{Password strength}\\\hline
Zxcvbn score & 2.20 & 2.14 & 0.75 & 0.67 & 4.00 \\
Shannon entropy & 29.31 & 28.37 & 7.93 & 16.00 & 68.00 \\
NIST entropy & 23.50 & 23.00 & 2.98 & 17.17 & 35.69 \\
\bottomrule
    \end{tabular}
\end{table}

As shown in Table~\ref{tab:em_breakdown}, the majority of the 1,767 logged passwords was entered with Chrome's auto-fill (53.71\%) followed by manual entry (33.39\%). Although in our pilot study various password manager plugins, e.g., KeePass and 1Password, had been correctly detected, in our actual study only LastPass was used by our participants (128 or 7.24\% of all passwords). Copy\&paste and unknown Chrome plugins formed the smallest, relevant-sized shares and only four passwords were entered programmatically by an external program.

\begin{table}[t]
    \caption{Number of distinct password entries with each entry method.}
    \label{tab:em_breakdown}
    \centering
    \begin{tabular}{l|r|r}
    \toprule
    \textbf{Entry method} & \textbf{All passwords} & \textbf{Unique passwords} \\\midrule
    Chrome auto-fill    & 949 (53.71\%) & 540 (51.67\%)\\
    Human               & 590 (33.39\%) & 331 (31.67\%)\\
    LastPass plugin     & 128 (7.24\%) & 100 (9.57\%)\\
    Copy\&paste          & 55 (3.11\%) & 51 (4.88\%)\\
    Unknown plugin      & 41  (2.32\%) & 23 (2.20\%)\\
    External manager    & 4   (0.23\%) & 0 (0.00\%)\\\midrule
    $\sum$              & 1,767          & 1,045\\
    \bottomrule
    \end{tabular}
\end{table}

\begin{figure}[t]
    \centering
    \includegraphics[width=\linewidth]{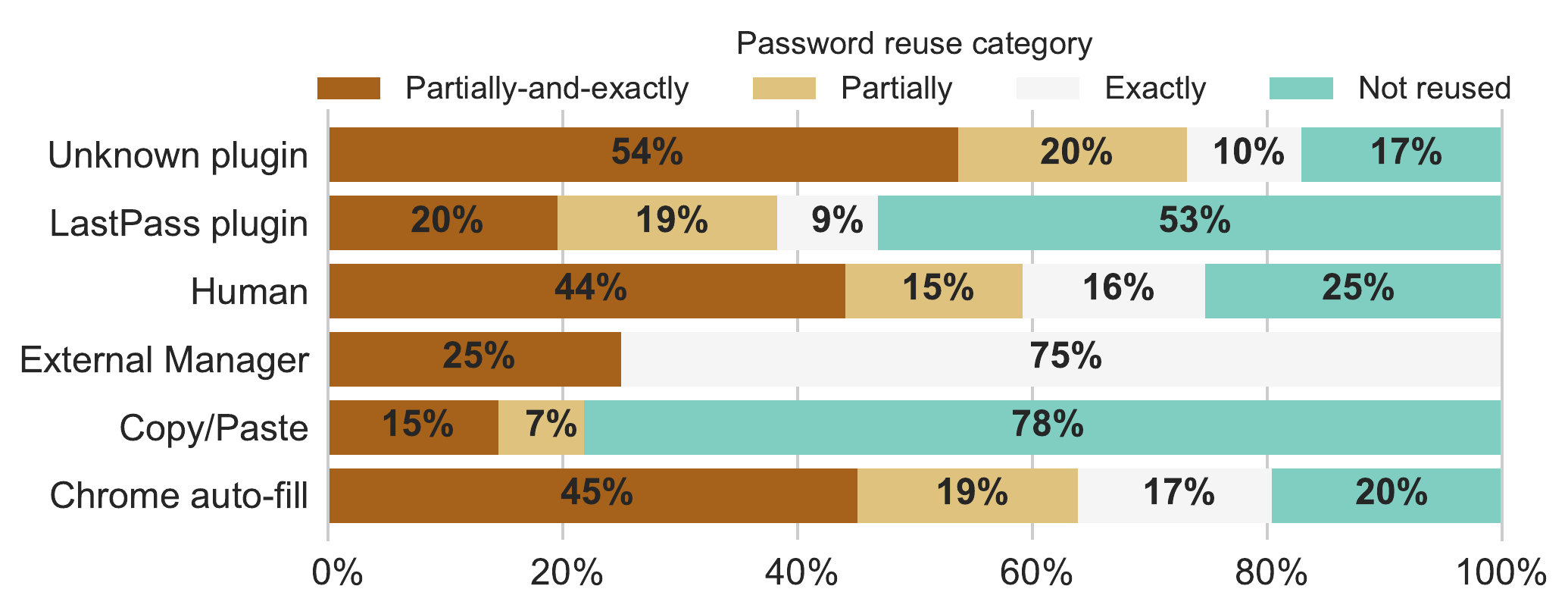}
    \caption{Breakdown of password reuse by entry method for all passwords.}
    \label{fig:em_reuse}
\end{figure}

\begin{figure}[t]
    \centering
    \includegraphics[width=\linewidth]{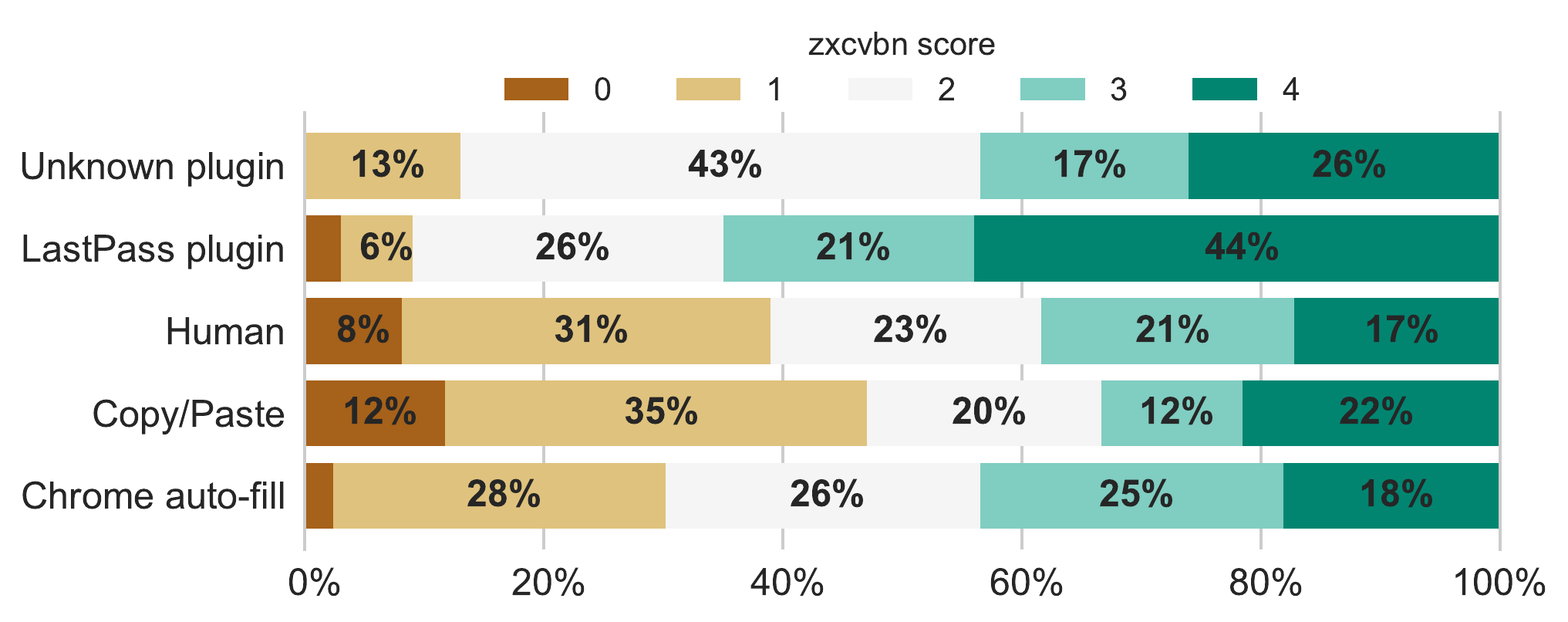}
    \caption{Breakdown of zxcvbn scores per entry method for unique passwords.}
    \label{fig:em_zxcvbn_breakdown}
\end{figure}

With respect to general password reuse (see Figure~\ref{fig:em_reuse}), partial-and-exact reuse is by far the most common reuse across all entry methods, except for LastPass' plugin and Copy\&paste, which have a noticeably high fraction of non-reused passwords (e.g., 68 or 53\% of all passwords entered with LastPass were not reused) and have noticeably less password reuse than the overall average. Looking at the password strength for all \textit{unique} passwords (see Figure~\ref{fig:em_zxcvbn_breakdown}), one can see that 65\% or 44 of all passwords entered with LastPass are stronger than the overall average of 2.20, while the other entry methods show a more balanced distribution across the zxcvbn scores (except for score 0). In summary, this indicates that LastPass shows an improved password strength (mean of 2.80 with SD=1.07) and password uniqueness in comparison to the other entry methods. Copy\&paste exhibits the strongest password uniqueness, however, at the same time the weakest password strength (1.98 on average with a SD=1.33).

\subsection{Grouping based on creation strategy}
We grouped our participants based on their self-reported strategies for creating new passwords (see \textit{Q9}, \textit{Q13}, and \textit{Q15} in Appendix~\ref{sec:questionnaire}).  Based on their answers, we discovered a dichotomous grouping: 

\subsubsection*{Group~1: Password managers/generators ("PWM")}

First, we identified participants that reported using a password generator, either as integrated part of a password manager program (e.g., \textit{"I use lastpass.com, which automatically creates and saves very strong passwords."}) or as an extra service (\textit{"I use a service to generate/create passwords that I put the parameters in that I would like."}). Many also implied the usage of a manager for password storage (e.g., \textit{"I use a password creation and storage-related browser extension that also is related to an installed password manager application on my personal computer."}), however, some participants explicitly noted a separate storage solution (\textit{"I use an app that creates random character strings to pick new passwords for me. I then memorize it so I don't have to keep it written down"} or \textit{"I will use a random password generator. [...] I will save the new password in a secure location such as a password protected flash drive."}). In total, 45 (or 26.47\%) out of 170 participants fit this category.

\begin{table}[t]
    \caption{Self-reported preferences for newly created passwords.}
    \label{tab:create_factor}
    \centering
    \begin{tabular}{l|c|c|c}
    \textbf{Group} & Ease of remembering & Security & Both \\\midrule
    Human   & 69 (57.02\%) & 51 (42.15\%) & 1 (0.83\%) \\\hline
    PWM     & 11 (22.45\%) & 38 (77.55\%) & 0  \\
    \end{tabular}
\end{table}

\subsubsection*{Group~2: Human-generated ("Human")}

We discovered that all 121 remaining participants described a strategy that abstains from using technical means (like password managers). Almost all of the participants in this group reported that they \textit{"try to come up with a (random) combination of numbers, letters, and characters,"} which prior work has shown to be prone to efficient statistical, data-driven attacks~\cite{6234435,Duermuth2015,197243,passgan}. For instance, one participant symptomatically reported: \textit{"I think of a word I want to use and will remember like. mouse. I then decide to capitalize a letter in it like mOuse. I then add a special character to the word like mOuse@. I then decided a few numbers to add like mOuse@84."} Only a very small subgroup of seven participants reported using analog tools to create passwords, such as dice or books (\textit{"I have a book on my desk I pick a random page number and I use the  first letter of the first ten words and put the page number at the end and a period after."}), or using passphrases (\textit{"i use song lyrics then add a random word at the end"}).

Many of the participants in this group also hinted in their answer to their password storage strategies. For instance, various participants emphasized ease of remembering as a criteria for new passwords (e.g., \textit{"something easy to remember, replace some letters with numbers."}; see also Table~\ref{tab:create_factor}), others use analog or digital storage (e.g., \textit{"I try to remember something easy or I right[sic] it down on my computer and copy\&paste it when needed."}). Many participants also outright admitted re-using passwords as part of their strategy (e.g., \textit{"I use the same password I always use because it has served me well all these years"} and \textit{"I have several go to words i use and add numbers and symbols that i can remember"}).

\begin{table}[t]
    \caption{Demographics of our two participant categories.}
    \label{tab:demographics_creation}
    \centering
    \scriptsize
    \begin{tabular}{l|r|r}\toprule
    & Human & PWM \\
    \hline
    \rowcolor[gray]{.9}
    \multicolumn{3}{l}{Number of participants}\\\hline
    & 121 & 49 \\\hline
    \rowcolor[gray]{.9}
    \multicolumn{3}{l}{Gender}\\\hline
    Female      & 59 (48.76\%)  & 14 (28.57\%) \\
    Male        & 62 (51.24\%)  & 35 (71.43\%)  \\
    \hline
    \rowcolor[gray]{.9}
    \multicolumn{3}{l}{Age group}\\\hline
    18--30      & 48 (39.67\%)  & 16 (32.65\%)         \\
    31--40      & 39 (32.23\%)  & 24 (48.98\%)  \\
    41--50      & 27 (22.31\%)  & 5 (10.20\%)  \\
    51--60      & 5  (4.13\%)   & 3 (6.12\%)          \\
    61--70      & 2  (1.65\%)   & 0           \\
    $\geq$71    & 0  0          & 1 (2.04\%) \\
    \hline
    \rowcolor[gray]{.9}
    \multicolumn{3}{l}{Computer science background}\\\hline
    Yes         & 10 (8.26\%)    & 17 (34.69\%) \\
    No          & 111 (91.74\%)   & 32 (65.13\%)  \\
    \hline
    \rowcolor[gray]{.9}
    \multicolumn{3}{l}{Education level}\\\hline
    Less than high school   & 0            & 1 (2.04\%) \\
    High school graduate    & 22 (18.18\%) & 4 (8.16\%)  \\
    Some college, no degree & 28 (23.14\%) & 6 (12.24\%) \\
    Associate's degree      & 27 (22.31\%) & 7 (14.29\%) \\
    Bachelor degree         & 35 (28.93\%) & 27 (55.10\%) \\
    Ph.D.                   & 0            & 1 (2.04\%) \\
    Graduate/prof. degree   & 9 (7.44\%)   & 3 (6.12\%)  \\
    \hline
    \rowcolor[gray]{.9}
    \multicolumn{3}{l}{Ethnicity}\\\hline
    White/Caucasian         & 91 (75.21\%)  & 32 (65.31\%) \\
    Black/African American  & 15 (12.40\%)  & 10 (20.41\%)   \\
    Asian                   & 5 (4.13\%)    & 4 (8.16\%)  \\
    Hispanic/Latino         & 10 (8.26\%)   & 2 (4.08\%) \\
    Multiracial             & 0             & 1 (2.04\%)  \\
    \hline
    \rowcolor[gray]{.9}
    \multicolumn{3}{l}{Privacy concern (Westin index)}\\\hline
    Privacy fanatic     & 45 (37.19\%)   & 21 (42.86\%)  \\
    Privacy unconcerned & 15 (12.40\%)   & 16 (32.65\%)          \\
    Privacy pragmatist  & 61 (50.41\%)   & 12 (24.49\%) \\
    \hline
    \rowcolor[gray]{.9}
    \multicolumn{3}{l}{Attitude about passwords}\\\hline
    Pessimist   & 1 (0.83\%)    & 2 (4.08\%)  \\
    Optimist    & 88 (72.73\%)  & 44 (89.80\%)             \\
    Conflicted  & 32 (26.45\%)  & 3 (6.12\%)  \\
    \hline
    \rowcolor[gray]{.9}
    \multicolumn{3}{l}{Prior password leaked experienced}\\\hline
    No              & 53 (43.80\%)   & 19 (38.78\%)  \\
    Yes             & 44 (36.36\%)   & 14 (28.57\%)          \\
    Not aware of    & 24 (19.83\%)   & 16 (32.65\%) \\
    \bottomrule
   \end{tabular}
\end{table}

\subsubsection{Group demographics}

We provide an overview of the groups' demographics in Table~\ref{tab:demographics_creation}. We again used a Mann-Whitney rank test to detect any significant differences in the distributions of those two demographic groups. We find that the two groups have statistically significant different distribution for gender ($U=2,366$, $p=.016$), computer science background ($U=2,181$, $p<.001$), and attitude towards passwords ($U=3,440$, $p=.024$). More participants in group~1 (PWM) identified themselves as male in comparison to group~2 (Human). The fractions of participants that have a computer science background and that are optimistic about passwords are higher in the group of password manager users. Gender and computer science background are significantly correlated for our participants (Fisher's exact test: $OR=3.99$, $p=.005$) as are computer science background and password attitude (chi-square test: $\chi^2=9.24$, $p<.01$). One hypothesis for this distribution could be that computer science studies had historically more male students and that their technical background may have induced awareness of the importance of passwords as a security measure and the promised benefits of password managers.

\begin{figure}
    \centering
    \includegraphics[width=\linewidth]{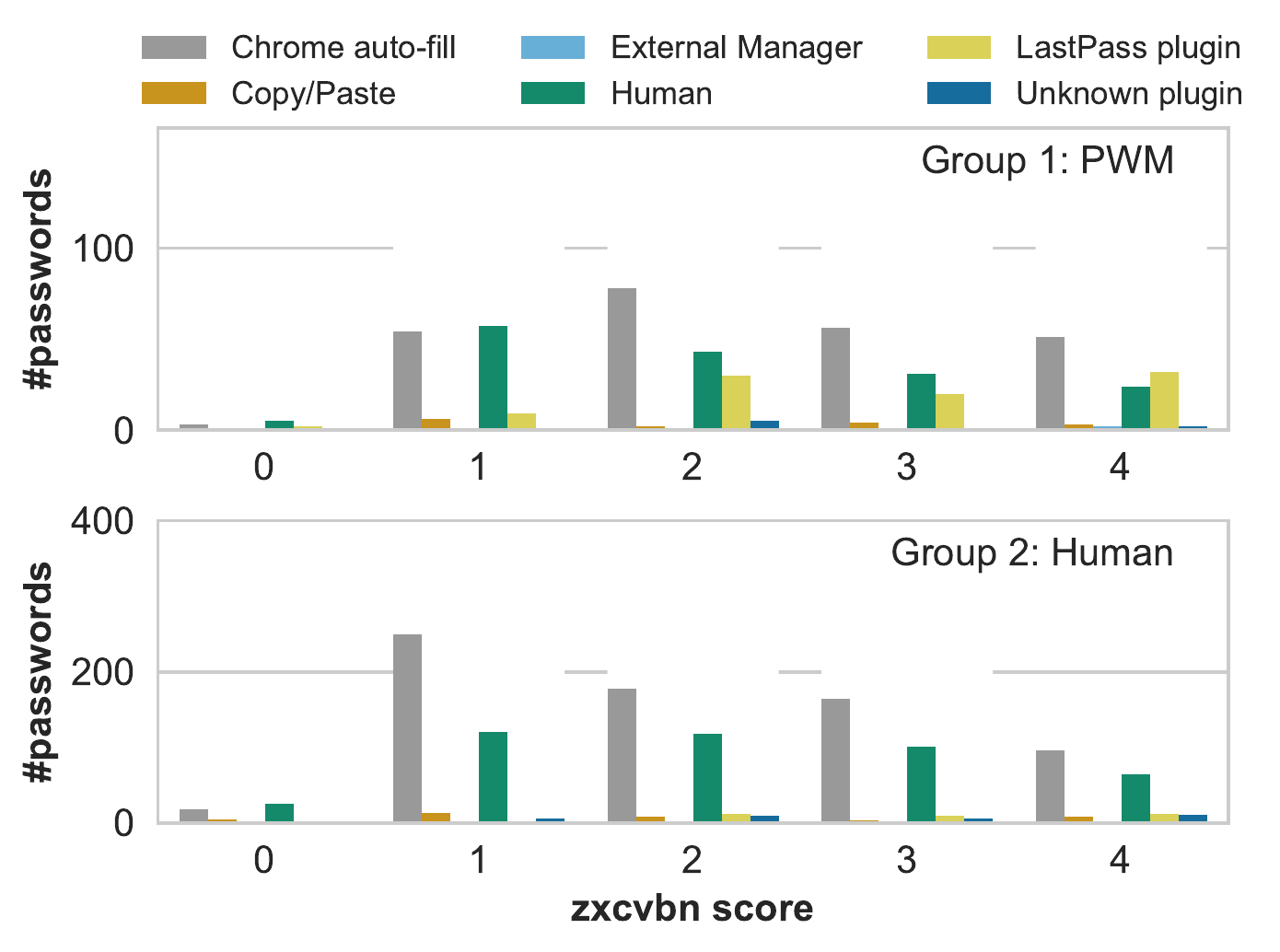}
    \caption{Password strength distribution by participant group and broken down by detected entry method. Hatched bars show total number of passwords per score. (Note the different y-axis limits)}
    \label{fig:em_zxcvbn_grouped}
\end{figure}

\begin{figure}
    \centering
    \includegraphics[width=\linewidth]{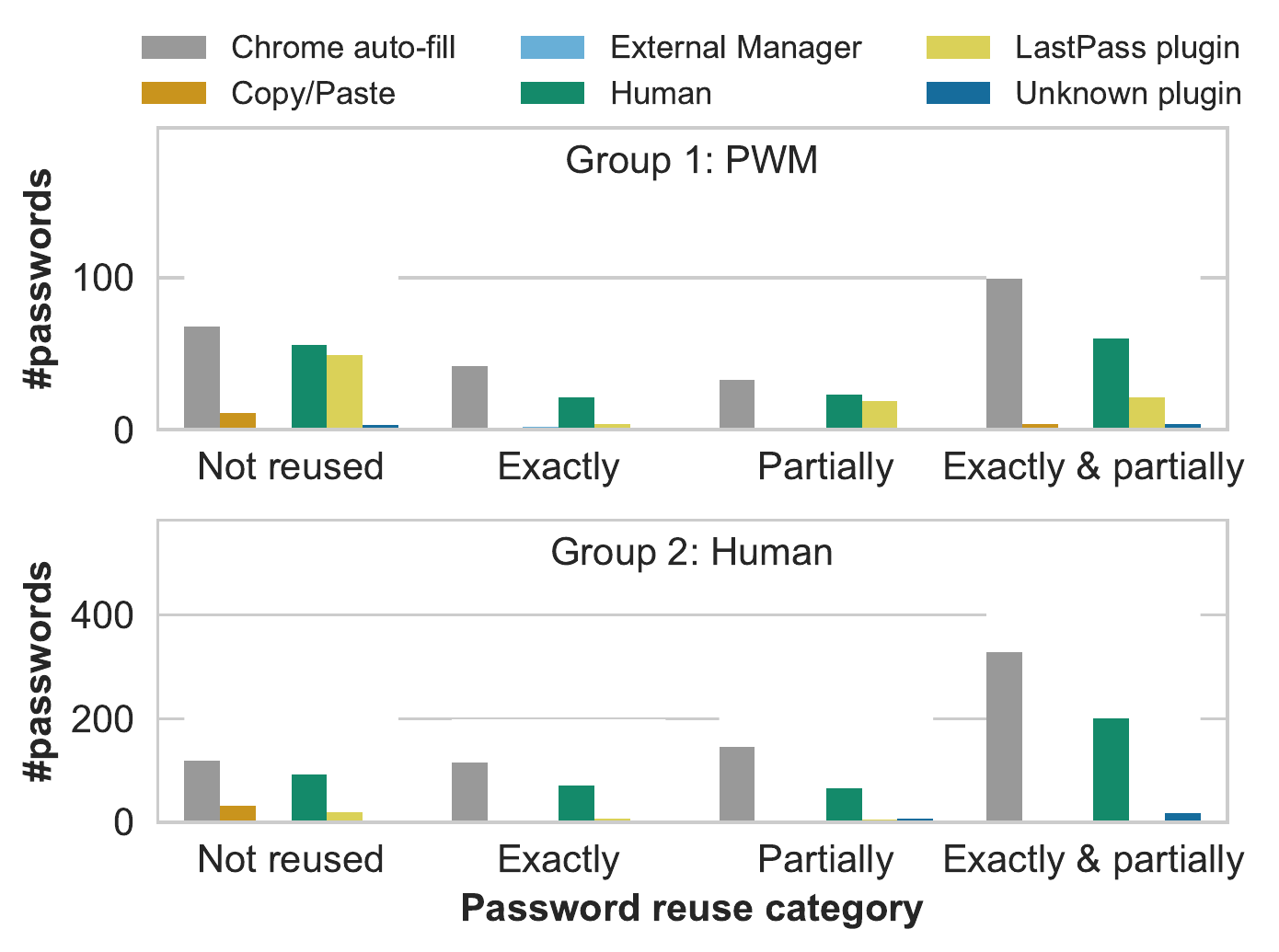}
    \caption{Distribution of password reuse categories by participant group and broken down by detected entry method. Hatched bars show total number of passwords per category. (Note the different y-axis limits)}
    \label{fig:em_reuse_grouped}
\end{figure}

\begin{table}[t]
    \caption{Distribution of entry methods per participant group.}
    \label{tab:em_per_group}
    \centering
    \begin{tabular}{lrr}
        \toprule
        \textbf{Entry method} &   \textbf{Group 1 (PWM)} &  \textbf{Group 2 (Human)} \\
        \hline
        \rowcolor[gray]{.9}
        \multicolumn{3}{l}{All passwords}\\\hline
        Chrome auto-fill &  242 (46.36\%) &  707 (56.79\%) \\
        Human            &  160 (30.65\%) &  430 (34.54\%) \\
        LastPass plugin  &   93 (17.82\%) &   35 (2.81\%)  \\
        Copy\&paste       &   16 (3.07\%)  &   39 (3.13\%)  \\
        Unknown plugin   &    8 (1.53\%)  &   33 (2.65\%)  \\
        External manager &    3 (0.57\%)  &    1 (0.08\%)  \\
        $\sum$           & 522 & 1245\\
        \hline
        \rowcolor[gray]{.9}
        \multicolumn{3}{l}{Unique passwords}\\\hline
        Chrome auto-fill &  144 (42.99\%)&  396 (55.77\%)\\
        Human            &  101 (30.15\%)&  230 (32.39\%)\\
        LastPass plugin  &   72 (21.49\%)&   28 (3.94\%) \\
        Copy\&paste       &   14 (4.18\%) &   37 (5.21\%) \\
        Unknown plugin   &    4 (1.19\%) &   19 (2.68\%) \\
        $\sum$           & 335 & 710\\
        \bottomrule
    \end{tabular}
\end{table}

\subsubsection{Comparison of password strength and reuse}

Figures~\ref{fig:em_zxcvbn_grouped} and \ref{fig:em_reuse_grouped} provide a comparison of the password strength and reuse between the two participant groups. The hatched bars indicate the overall number of passwords per zxcvbn score and reuse category, respectively. The plain bars break the number of passwords down by entry method. Participants in group 1 (PWM) entered in total 522 passwords and participants in group 2 (Human) entered in total 1245 passwords (both numbers include reused passwords, see Table~\ref{tab:em_per_group}).

For password strength (see Figure~\ref{fig:em_zxcvbn_grouped}), neither group contained a noticeable fraction of the weakest passwords with score 0. However, group~2 shows a clear tendency towards weaker passwords. For instance, there are almost twice as many score~1 passwords ($n=390$) than score~4 passwords ($n=191$). In contrast, the most frequent score for group~1 is 2 ($n=158$), but the distribution shows a lower kurtosis (e.g., scores 1,3, and 4 have the frequencies 126, 113, and 114). When breaking the number of passwords down by their entry method, Chrome auto-fill is the dominating entry method for all zxcvbn scores 1--4 in both groups except for score 1 in group~1, where manually entered passwords are most frequent. However, for group~1 the fraction of passwords entered with LastPass' plugin ($n=93$ or 17.82\% of the passwords) is considerably larger than for group~2 ($n=35$ or 2.81\%). In particular, for group~1, passwords entered with LastPass have mostly scores higher than 2 ($n=82$), where score~4 is the most frequent ($n=32$).

Regarding password reuse (see Figure~\ref{fig:em_reuse_grouped}), the most frequent category is exactly-and-partially reused ($n=189$ or 36.21\% for group~1; $n=555$ or 44.58\% for group~2). However, group~1 shows a bimodal distribution in which not-reused passwords are almost as frequent ($n=187$) as exactly-and-partially reused ones. Further, Chrome auto-fill is the dominating entry method for all reuse categories in both groups. However, when breaking the passwords down by entry method, more than half ($n=49$ or 52.69\%) of the passwords entered with LastPass in group~1 have not been reused in any way. The vast majority of reused passwords can be attributed to manual entry and Chrome auto-fill. In group~1, 335 or 64.18\% of the passwords have been reused and in group~2 979 or 78.63\% of the passwords. Of the 335 reused passwords in group~1, 278 or 82.99\% passwords have been entered manually or with Chrome auto-fill. In group~2, 926 or 74.38\% of the reused passwords were entered manually or with Chrome auto-fill.

\subsection{Case-study: Active LastPass users}
\label{sec:casestudy}

\begin{figure}[t]
    \centering
    \includegraphics[width=\linewidth]{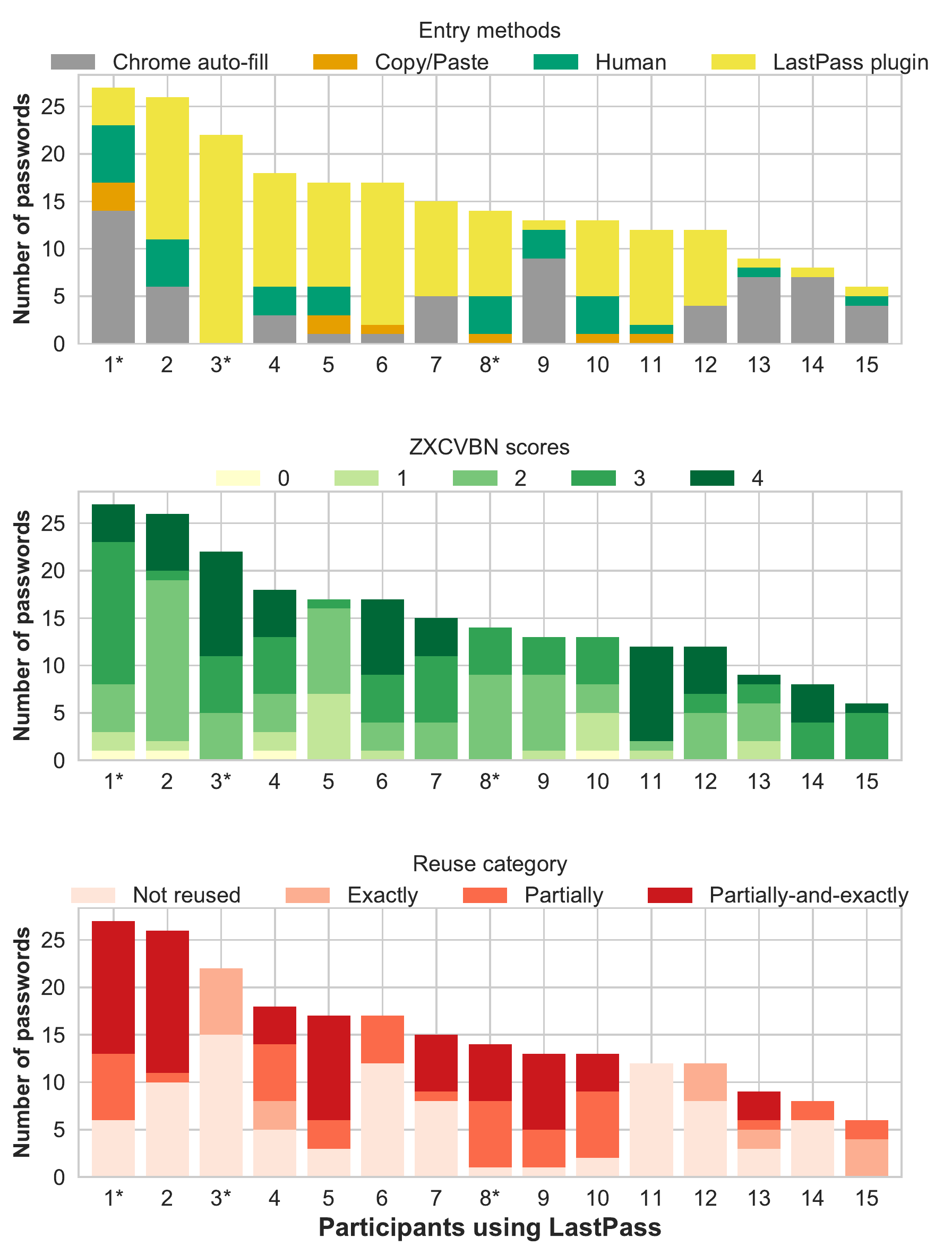}
    \caption{Distribution of entry methods, zxcvbn scores, and password reuse for active LastPass users. Users from group~2 (Human) are marked *.}
    \label{fig:lpworkers}
\end{figure}

We were interested in how consistent users of password managers employ their tools during their normal web browsing. Our dataset contains 15 users that entered at least one password with a known password manager plugin and for which we are hence certain that they are users of a password manager solution (e.g., we cannot be certain about users that copy\&paste all their passwords from a manager into the password fields). All of those 15 users employ LastPass as manager. Figure~\ref{fig:lpworkers} gives an overview of those 15 users' password properties.

We can observe that, except for user~3, all users entered passwords through at least one additional entry method, most even two methods, however, LastPass' plugin is the primary entry method for 10 of the 15 users and on average 52\% of the passwords in this selection were entered through LastPass (SD=31\%). Interestingly, user~3 gave no indication in the survey sampling for using a password manager and hence is in group~2 (Human), but was the only user to enter all passwords through LastPass. Of the 15 users, four users did not enter any strong password with zxcvbn score 4 and every user entered at least one password with zxcvbn score smaller than 4 through LastPass. Nevertheless, the mean zxcvbn score (mean=2.72, SD=0.58) of this selection of participants is above the global average. All but user~11 reused at least one password either partially, exactly, or both. User~15 even reused all of their passwords. The average user in this selection reused 60\% of their passwords. That is below the global average in our dataset.

Although there are some users that seem to particularly benefit from using LastPass (e.g., users 3, 6, 11, and 12 are heavy LastPass users with low reuse and strong passwords), we could neither confirm nor refute a statistically significant correlation between ratio of LastPass passwords per user and either ratio of strong or non-reused passwords, since presuming a small or medium sized effect the number of LastPass users in our dataset is too small for a statistical test with sufficient testpower.

\subsection{Modelling password strength and reuse}

To get a better understanding of the influencing factors for password strength and reuse we conducted several regression analyses that include the effects of the users strategies as well as their password manager usage. To account for the hierarchical structure of our data, where individual password entries are grouped under the corresponding participant, we calculated multi-level (aka hierarchical) logistic- or ordinal-regressions that allow the intercepts to vary at the participant level. By comparing simple and multi-level models for reuse and strength, the significant superiority of the latter was demonstrated. Thus, we report here only the final multi-level regression models and how they were constructed.

\subsubsection{Correlation analysis}

Before constructing the models, we started out with a correlation analysis of the available factors (e.g., password composition, participant group, self-reported website value, etc.). As multi-level models are highly vulnerable to multicollinearity, detecting and potentially removing strongly correlated variables is essential to prevent inaccurate model estimations which could lead to false positive results. In our dataset, we detected a very high, significant correlation between zxcvbn scores and password composition, in particular password length, as well as with the NIST and Shannon entropies. Since we consider zxcvbn a more realistic measurement of crackability, we omitted NIST and Shannon entropies from our model. Investigation of zxcvbn showed that zxcvbn rewards lengthy passwords with better scores and that its pattern and l33t speak detection can penalize passwords with digits and special characters. Since zxcvbn is the more interesting factor for us and since it partially contains the effect of the password composition on the prediction, we excluded password composition parameters from our models. Moreover, we noticed that password reuse was strongly correlated with the presence of a lowercase character in the password. A closer inspection of our dataset showed, that our data contained a number of PINs, which were all unique, and that every non-PIN password contains at least one lowercase character. In this situation, including the presence/absence of lowercase characters would result in our model just distinguishing between PINs and non-PINs when predicting password reuse.

\paragraph{Website category as proxy for website value}
 
\begin{figure}[t]
     \centering
     \includegraphics[width=\linewidth]{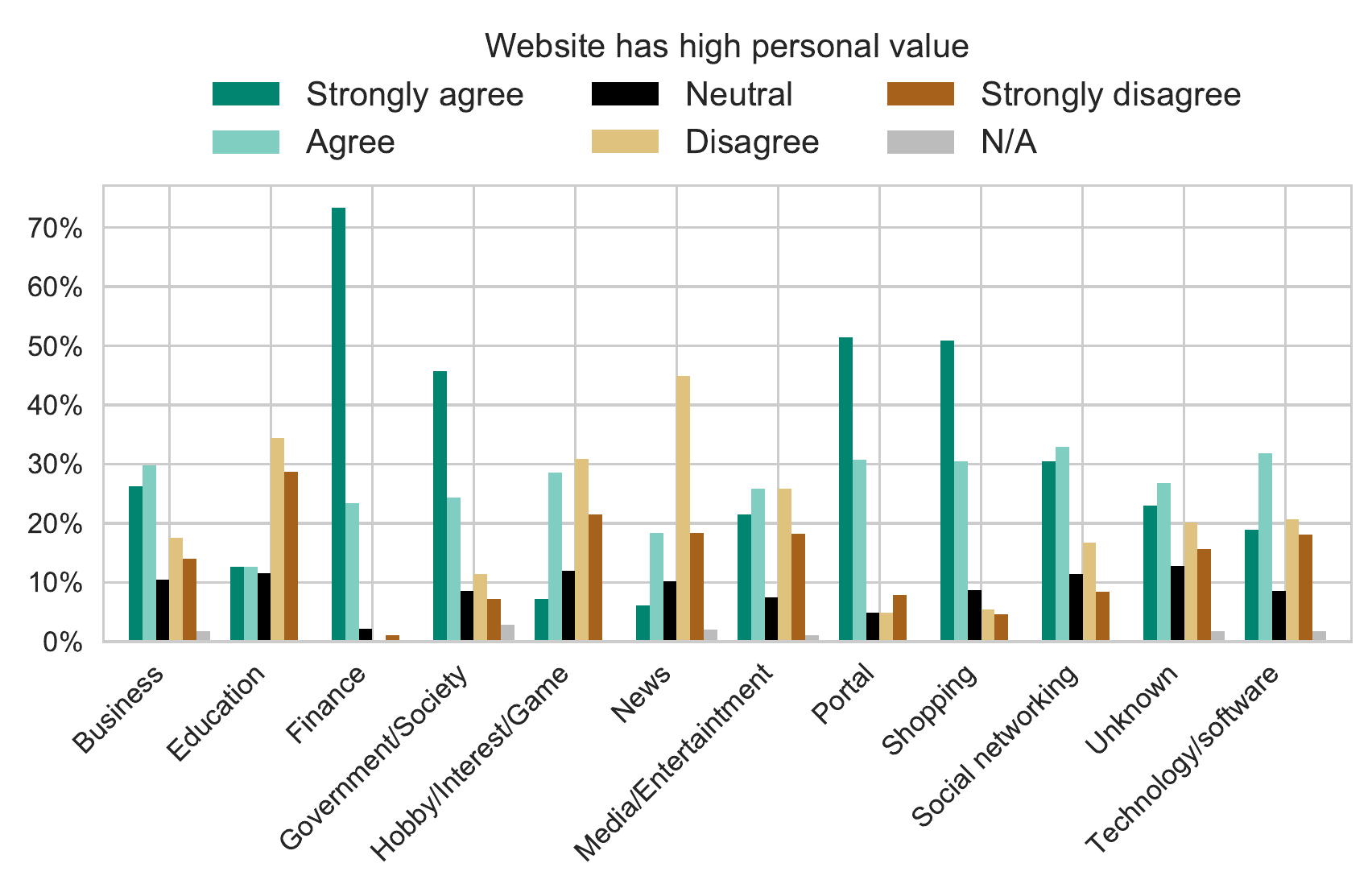}
     \caption{Self-reported website value per website category}
     \label{fig:cat_value_map}
 \end{figure}

Commonly the website category is used as a proxy for the website value. Since we collected both, self-reported website value from the in-situ questionnaire and website category from the domain, we can provide insights into this general assumption. Figure~\ref{fig:cat_value_map} shows the self-reported value per respective domain. For instance, in more than 70\% of all logged passwords for a financial domain, the user reported a very high value for that domain. Similarly, in more than 60\% of all logged passwords for news websites, the users (strongly) disagreed that this domain has a high value. Unfortunately, domains with an unknown category, which form the bulk of our logged passwords (632 or 35.8\% of all logged entries; see Table~\ref{tab:websitecategories}), did not show a clear tendency towards high or low value.

Although prior works~\cite{naturalhabitat} used the website category in comparable models, in our regression models, we decided for above stated reasons on using the self-reported website value instead of the website category as a predictor.
 
  \begin{table}[t]
    \caption{Number of logins for each website category by our participants.}
    \label{tab:websitecategories}
    \centering
    \begin{tabular}{l|c|c|}
    \toprule
    \textbf{Category} & \textbf{Count} & \textbf{\%} \\\midrule
    Shopping &                             240 & 13.9\\
    Social networking &                    167 & 9.5\\
    Tech/Software/Filesharing &            116 & 6.6\\
    Portal &                               101 & 5.7\\
    Finance &                               94 & 5.3\\
    News/Media/Entertainment &              93 & 5.3\\
    Education &                             87 & 4.9\\
    Government/society &                    70 & 4.0\\
    Business \& economy &                    57 & 3.2\\
    News/Headline\_Links/Collaborative &     49 & 2.8\\
    Hobby/Interest/Game &                   42 & 2.4\\
    Health &                                 6 & 0.3\\
    Insurance/Auto &                         6 & 0.3\\
    Job/search &                             4 & 0.2\\
    Research Methods/Science &               2 & 0.1\\
    Resources &                              1 & 0.1\\
    Unknown &                              632 & 35.8\\
    \bottomrule
    \end{tabular}
\end{table}

\begin{table}[t]
    \caption{Goodness of fit for the models predicting ZCVBN scores}
    \label{tab:anova_zxcvbn}
    \centering
    \footnotesize
    \begin{tabular}{lrrrr}\toprule
                            & AIC      & logLik     & df    & Pr(>Chisq) \\\midrule
    simple regression       & 5080.6   & -2536.3\\   
    \rowcolor[gray]{.9}
    multi-level base      & 4536.7   & -2263.4    & 1     & <0.001\\
    \rowcolor[gray]{.9}
     + login level         & 4316.3   & -2147.1    & 6     & <0.001\\
     + user level      & 4320.4   & -2143.2    & 6     &  0.2494034\\
    \rowcolor[gray]{.9}
     + interactions  & 4309.5   & -2133.7    & 4     &  <0.001\\
    \bottomrule
    \end{tabular}
\end{table}

\begin{table}[t]
    \caption{Goodness of fit for the models predicting password reuse}
    \label{tab:anova_reuse}
    \centering
    \footnotesize
    \begin{tabular}{lrrrr}\toprule
                             & AIC    & logLik   & Df   & Pr(>Chisq)  \\\midrule
    simple regression  & 1959.7 & -978.84  \\    
    \rowcolor[gray]{.9}
    multi-level base       & 1794.6 & -895.28  & 1    & < 0.001\\    
    \rowcolor[gray]{.9}
    + login level        & 1694.9 & -839.46  & 6    & < 0.001\\
    \rowcolor[gray]{.9}
    + user level     & 1684.7 & -828.37  & 6   & <0.01\\
    + interactions  & 1687.6 & -825.80  & 4   & 0.27351\\
    \bottomrule
    \end{tabular}
\end{table}

\subsubsection{Constructing the models}

Selecting an appropriate model corresponding to the empirical data is a crucial step in every regression based analysis. This process ensures that only sets of variable are included that significantly explain variation in the empirical data. To this end, for both password reuse and password strength prediction, we started with a base model without any explanatory variables, which we iteratively extended with additional predictors. Tables~\ref{tab:anova_zxcvbn} and \ref{tab:anova_reuse} present the goodness of fit for the relevant steps in this model building process. According to the scale level of our dependent variable and the hierarchical structure of our data, we build an ordinal multi-level regression model for the zxcvbn scores (Table~\ref{tab:anova_zxcvbn}) and a logistic multi-level regression model for the password reuse (Table~\ref{tab:anova_reuse}). To verify that a multi-level approach suits our data better than a simple regression model we first tested basic multi-level models for password reuse and password strength without any explanatory variables against the corresponding simple regression models. Both multi-level models fitted the data significantly better.

\paragraph*{Process of model fitting}
Throughout the main process of model selection, we extended both multi-level models in three steps by adding sets of predictors. In three steps we included: 1)~variables measured at the \textit{login level}: a)~the entry method of the password, b)~the self-reported website value and c)~the self-reported password strength; 2)~variables measured at the \textit{user level}: a)~the number of submitted passwords per user, b)~the password creation strategy of the user and c)~the self-reported password management strategies of the user; 3)~the \textit{cross level} interactions between the user’s password creation strategy an the detected entry method.

This approach not only allows us to evaluate the effects of the individual explanatory variables, but also to investigate the interplay between different storage strategies and the password creation strategy of the users. In each iteration we computed the model fit and used log likelihood model fit comparison to check whether the new, more complex model fit the data significantly better than the previous one. As our final model we picked the one with the best fit that was significant better in explaining the empirical data than the previous models. This is a well established procedure for model building in, e.g., social sciences and psychological research~\cite{hoxbook,discoverstatistics,lme4_r,ordinal_r}, and allows creation of models that have the best trade-off of complexity, stability, and fitness.

\paragraph*{Selecting the appropriate model}
All models are compared according to the corresponding akaike information criterion~(AIC), which is an estimator of the relative quality of statistical models for a given set of data. Smaller AIC scores indicate a better fitting model. Additionally, the models are statistically compared using likelihood-ratio tests, which are evaluated using a Chi-squared distribution. The final model is selected based on AIC as well as their ability to describe the empirical data better than the previous models.

\begin{table}[t]
    \caption{Logistic multi-level regression model predicting zxcvbn score. Estimates are in relation to manually entered passwords by a human. Statistically significant predictors are shaded. Interactions are marked with *.}
    \label{tab:model_zxcvbn}
    \centering
    \scriptsize
    \begin{tabular}{rrrrr}\toprule
    & Estimate & Std. Error & z value & Pr(>|z|)\\\midrule
    em:chrome                       & 0.07  & 0.12  & 0.59  & 0.56\\    
    em:copy/paste                   & -0.13 & 0.35  & -0.89 & 0.37\\
    em:lastpass                     & 0.24  & 0.35  & 0.69  & 0.49\\
    \rowcolor[gray]{.9}
    em:unknownplugin                & 1.02  & 0.34  & 2.97  & <0.01\\
    in-situ:value                   & 0.02  & 0.05  & 0.48  & 0.63\\
    \rowcolor[gray]{.9}
    in-situ:strength                & 0.89  & 0.07  & 12.68 & <0.001\\
    user:entries                    & 0.02 & 0.02  & 0.69  & 0.49 \\
    q9:generator                        & -0.45 & 0.67  &-0.68  & 0.50 \\   
    q14:memorize                    & -0.24 & 0.30  & -0.79 & 0.43\\
    q14:analog                      & 0.05  & 0.29  & 0.16  & 0.88\\
    q14:digital                     & 0.09  & 0.31  & 0.29  & 0.77\\
     q14:pwm                        & -0.16 & 0.28  & -0.57 & 0.57  \\
     \rowcolor[gray]{.9}
    em:chrome * q9:generator        & 2.30  & 0.60  & 3.84  & <0.001 \\
    \rowcolor[gray]{.9}
    em:copy/paste * q9:generator    & 3.40  & 1.22  & 2.79  & <0.01\\
    \rowcolor[gray]{.9}
    em:lastpass * q9:generator      & 1.83  & 0.82  & 2.24  & <0.05\\ 
    em:unknownplugin * q9:generator & 0.22  & 1.34  & 0.16  & 0.87\\\hline
    \multicolumn{5}{c}{{\tiny em: Entry method; q9: Creation strategy; q14: Storage strategy; in-situ: Plugin questionnaire}}\\
    \bottomrule
    \end{tabular}
\end{table}

\subsubsection{Model for zxcvbn score}
For the zxcvbn score an ordinal model with all predictors and also the mentioned interaction described our data best. The model is presented in Table~\ref{tab:model_zxcvbn}.

The interactions between the self-reported password creation strategy (\textit{q9:generator}; see \textit{Q9} in Appendix~\ref{sec:questionnaire}) and the detected entry methods Chrome auto-fill, copy\&paste, and LastPass were significant predictors in our model. Those entry methods and also the creation strategy are not significant predictors of password strength on their own. This means that using such a password management/entry tool only leads to significant improvement in the password strength when the users also employ some supporting techniques (password generator) for the creation of their passwords.

The model might suggest that a general password entry with a plugin (other than LastPass in our dataset) increased the likelihood of a strong password. However, this could be attributed to the high standard error resulting from the minimal data for this entry method.

Moreover the self-reported password strength was a significant predictor of the measured password strength. This indicates that the users may have a very clear view on the strength of the passwords they have entered.

\subsubsection{Model for password reuse}
For password reuse a logistical model with all predictors but without interactions described our data best. Table~\ref{tab:model_reuse} presents our regression model to predict password reuse. Reuse was significantly influenced by the entry method of the password. In contrast to human entry the odds for reuse were 2.85 time lower if the password was entered with LastPass (odds ratio 0.35, averaged predicted probability of reuse with Lastpass = 48,35\%) and even 14.29 times lower if entered via copy\&paste (odds ratio 0.07, averaged predicted probability of reuse with copy\&paste = 19,81\%). Interestingly, the input via Google Chrome auto-fill even had a negative effect on the uniqueness of the passwords. In contrast to human entry the odds for reuse were 1.65 times higher if the password was entered with Chrome auto-fill (odds ratio 0.35, averaged predicted probability of reuse with Chrome auto-fill = 83.72\%). A further significant predictor of password reuse is the user's approach of creating passwords. For users who use technical tools to create their passwords (\textit{q9:generator}), the chances are 3.70 times as high that the passwords are not reused (odds ratio 0.27, predicted probability of reuse if technical tools are used = 47,36\%). In contrast to the models explaining the zxcvbn-score, our data does not indicate the presence of an interaction effect of the user's password creation strategy on the relation between entry method and password reuse.

In addition, we found a positive relation between the the numbers of passwords entered by users and the reuse of these passwords. In our model, each additional password of the user increases the chance that it will be reused by 6\% (odds ratio 1.06). This suggests that with increasing numbers of passwords, it becomes more likely that some of them will be reused, which is in line with prior research~\cite{Gaw:2006:PMS:1143120.1143127}.

We also found the self-reported value and password strength of users a statistically significant predictor for reuse~\cite{DBLP:conf/scn/BaileyDP14}. Passwords entered to a website with a higher value for the user were less likely to be reused (odds ratio of 0.87) and also passwords that the users considered stronger were less likely to be reused (odds ratio of 0.81).

Lastly, users that reported using an analog password storage (\textit{q14:analog}; see \textit{Q14} in Appendix~\ref{sec:questionnaire}) were less likely to reuse their passwords (odds ratio of 0.62).

\begin{table}[t]
    \caption{Logistic multi-level regression model predicting password reuse. Estimates are in relation to manually entered passwords by a human. Statistically significant predictors are shaded.}
    \label{tab:model_reuse}
    \centering
    \scriptsize
    \begin{tabular}{rrrrr}\toprule
    & Estimate & Std. Error & z value & Pr(>|z|)\\\midrule
    (Intercept)                 & 2.62   & 0.45   & 5.80  & <0.001\\
    \rowcolor[gray]{.9}
    em:chrome                   & 0.46   & 0.16   & 2.81  & <0.01\\
    \rowcolor[gray]{.9}
    em:copy/paste               & -2.68  & 0.41   & -6.54 & <0.001\\
    \rowcolor[gray]{.9}
    em:lastpass                 & -1.05  & 0.37   & -2.86 & <0.01\\
    em:unknownplugin            & 0.76   & 0.51   & 1.51  & 0.13\\
    \rowcolor[gray]{.9}
    in-situ:value               & -0.13  & 0.06   & -2.01 & <0.05  \\ 
    \rowcolor[gray]{.9}
    in-situ:strength            & -0.21  & 0.08   & -2.50 & <0.05  \\
    \rowcolor[gray]{.9}
    user:entries                & 0.06   & 0.02   & 2.67  & <0.01 \\ 
    \rowcolor[gray]{.9}
    q9:generator                & -1.31  & 0.40   & -3.24 & <0.01\\ 
    q14:memorize                & 0.22   & 0.25   & 0.88  & 0.38 \\
    \rowcolor[gray]{.9}
    q14:analog                  & -0.48  & 0.24   & -1.98 & <0.05\\
    q14:digital                 & -0.18  & 0.26   & -0.70 & 0.48\\
    q14:pwm                     & -0.07  & 0.24   & -0.30 & 0.76\\\hline
    \multicolumn{5}{c}{{\tiny em: Entry method; q9: Creation strategy; q14: Storage strategy; in-situ: Plugin questionnaire}}\\
    \bottomrule
    \end{tabular}
\end{table}

%% file: sections/discussion.tex
\section{Discussion}
\label{sec:discussion}

We discuss the results and limitations of our study on password managers' impact on password strength and reuse.

\subsection{Password Managers' Impact}

In general, our participants showed very similar password strength and reuse characteristics as in prior studies~\cite{naturalhabitat,197316} and our analysis could also reaffirm prior results, such as rampant password reuse and high share of low-strength passwords, and extend them, e.g., when asked in-situ users made very accurate estimates of their passwords' strength.

Our study adds novel insights to the existing literature by considering the exact password entry methods and when painting a more complete picture by considering the users' password creation strategies (e.g., consider the following participant statement for a mixed strategy: \textit{"I think Keepass is 100\% safe but it is annoying having to copy and paste passwords every time so I also use Chrome's built in password saving feature."}). We found that almost all users entered passwords with more than one entry method. Further, we discovered that every entry method exhibited reused passwords, although the ratio of reused passwords differ significantly between the entry methods. More than 80\% of Chrome auto-filled passwords were reused, while only 47\% of the passwords entered with LastPass' plugins were reused in some way, and even only 22\% of the copy/pasted passwords. Similarly, we noticed that low-strength passwords have been entered with all entry methods, where LastPass had on average the strongest passwords (mean of 2.80). Interestingly, manually entered passwords and Chrome auto-filled passwords showed very similar characteristics of being on a par with the overall password strength, but showing above average reuse rate.

When looking at active users of 3rd party password management software in separation, we found that only very few users came close to the picture of an "ideal" password manager user, i.e., all passwords entered with the manager, no password reuse, exclusively strong passwords. Instead, most of those users show a mixture of entry methods, password strength, and password reuse. Perhaps interestingly, not copy/paste, as might be expected from password managers' workflows, but Google's auto-fill feature was the second most used entry method after LastPass' plugin. This highlights, that to understand the influence of password managers, not only the entry methods, but also the users' strategy has to be taken into account.

For our participants, we discovered a dichotomous distribution of self-reported creation strategies. Participants indicated using a password generator right now or in the recent past, or clearly described mental algorithms and similar methods for human-generated passwords; only a negligible fraction of participants mentioned analog tools or alternative strategies (like two-factor authentication). Taking a differentiated view based on the creation strategies, we find that users of a password generator are closer to a desirable situation with stronger, less reused passwords, although being far from ideal.

Using regression modelling, we put our data together to a more complete view of password manager's influence. Our models suggest that the interaction between the creation strategy and the entry methods has significant influence on the password strength. If the passwords are entered with technical support (auto-fill, password manager plugin, or copy\&paste), this results in stronger passwords under the condition that technical means were already used when generating the passwords in the first place. Thus, password managers that provide users with password creation features indeed positively influence the overall password strength in the ecosystem. All the more, it is curious that Google Chrome, as the primary tool to access websites, has the password generation feature disabled by default~\cite{enable_chrome_generator}. Future work could investigate and compare Apple's walled-garden ecosystem, where the Safari browser has this feature enabled by default.

Our models further suggest, that the use of password generators and the website value also significantly reduced the chance of password reuse. More interestingly, however, is that the password storage strategies have different influence independently of an interaction with the creation strategy. Using a password manager plugin or copy\&pasting passwords reduced password reuse, while Chrome's auto-fill aggravated reuse. In other words, we observed that users were able to \textit{manually} create more unique passwords when managing their passwords digitally or with a manager, but not with Chrome auto-fill.

The benefit of password managers is also put into better perspective when considering particular strategies in our group~2 (human-generated passwords). We noticed that users tend to have a "self-centered" view when it comes to passwords' uniqueness (i.e., personal vs.~global), but are unaware of the fact that an attacker would not be concerned with personal uniqueness of passwords. A large fraction of users reported to \textit{"come up with [a password they] have never used before"}, to \textit{"use some words [they] would be familiar with but other weren't"}, or to \textit{"try to think of something that [they] have never used before"}. Those results also align with prior studies of password behavior~\cite{185315,rinn2015password,Inglesant:2010:TCU:1753326.1753384}.

Further, in light of the high relevance of copy\&paste for strong and unique passwords, our results can also underline the "Cobra effect"~\cite{nopaste1,nopaste2,nopaste3} of disabling paste functionality for password fields on websites to encourage the use of two-factor authentication or usage of password managers. 

The conclusion we draw from our results is that password managers indeed provide benefits to the users' password strength and uniqueness. Although both benefits can be achieved separately, our data suggests that indeed the integrated workflow of 3rd party password managers for generation and storage provides the highest benefits. However, we did not detect the \textit{ideal} user of password managers, but, on the opposite, that even users of 3rd party password manager still exhibit a high ratio of reused passwords and less than desired strong passwords. Even more troublesome is that our results suggest that the most widely used manager, Chrome's password saving and auto-filling feature, has only a positive effect on password strength when used in conjunction with an additional generator and even has an aggravating effect on password reuse. For studying password managers' influence on password security, our results show that a more holistic view has to be taken that considers both the exact password entry method and users' password creation and storage strategy. Simply focusing on only one of those factors will not yield the necessary insights.

\subsection{Impressions of password managers}

Lastly, we collected from our participants at different points in our survey---survey sampling and exit survey---their impressions and opinions about password managers. Those collected information provide insights into why users abstain from using password managers, to which contexts users restrict usage of managers, and also hint at misunderstood security benefits of managers.

\subsubsection*{Reasons for abstaining from password managers}

Given the benefits of password managers, we were interested in our participants' reasons to abstain from using them. Table~\ref{tab:exit_questionnaire} summarizes the results of our exit survey. We found that users have a discomfort with relinquishing control of their passwords to password managers, they do not believe that password managers provided more security, or see a non-necessity of password managers. Noticeably, the same reasons were given not only for abstaining from password managers, but also why users stopped using them. Of our 109 exit survey participants, 19 affirmed stopping to use a password manager for one of said reasons. Prototypical answers were \textit{"It was confusing, I did not know how to use it, and was afraid of being blocked out of my passwords."}, \textit{"I became skeptical of its security. I did a little research and found its possible to crack"}, or \textit{"After the free trial was up they wanted a fee, so I cancelled it.  I felt that it was not that useful to me. The browser seemed to do the same thing with more efficiency and less bother.  Plus the program was annoying .I kept offering a pop up asking me for this and that."} The answers we received from our participants in our exit survey (\textit{ES1} and \textit{ES4} in Appendix~\ref{sec:exitsurvey}) are aligned with the results of prior studies~\cite{Chiasson:2006:USC:1267336.1267337,Karole:2010:CUE:2041036.2041056}.

\begin{table}[t]
    \caption{Exit Survey's Result}
    \label{tab:exit_questionnaire}
    \centering
    \begin{tabular}{l|r|r}\hline
    \rowcolor[gray]{.9}
    \multicolumn{2}{l}{Users do not use any kind of 3rd party password managers because...}\\\hline
    the participants do not trust vendor/software &  37.61\% \\
    they cost lots of money &  20.18\% \\
    they are not really easy to set up/ easy to use &  11.93\% \\
    of lack of synchronization between users' different device & 7.34\%  \\
    of lack of support for the user device & 2.75\% \\
    \hline
    \rowcolor[gray]{.9}
    \multicolumn{2}{l}{Other Reasons}\\\hline
    Chrome's password saving feature suffices for the users &  59.63\% \\
    can handle managing the password w/o manager & 37.61\%  \\
    did not think about it before &  29.36\% \\
    not sure which one is better &  24.77\%\\
    \bottomrule
    \end{tabular}
\end{table}

\subsubsection*{Password managers as single point of failure}

We also noticed the high amount of users' distrust into password managers in our participants' answers in our survey sampling. When asked about their impressions' of password managers, various participants expressed concerns about managed passwords as a single point of failure. The concerns included software security issues, such as
\begin{itemize}
    \item \textit{"I think that it saves time but also generates a way for hackers to steal the information for themselves."}
    \item \textit{"It's not that secure, if someone managed to hack me or get a virus in they'd get everything stored in there."}
    \item \textit{"I would never use my browser's manager should my computer ever be hacked. I do trust LastPass for my pass word protection and do change my passwords on a regular basis. I keep my LastPass main password written down in a secure place in my home."}
    \item \textit{"I can see using the password saving feature of a browser as being convenient, but it leaves the user vulnerable to hackers who find a security breach."}
    \item \textit{"I feel that using my browser password saving feature is dangerous so I hate to think about it. If hackers hack just my browser they would have a bunch of passwords. I don't store them all on there for that reason."}
    \item \textit{"I don't like using the browser saving feature, makes me feel like someone can hack into my browser, like Google and then get to my passwords.  I do use it though for non-security sites, like signing up for emails from an on-line store, but not for anything that needs stronger security. I like some of the password managers, but I feel like I am not utilizing them efficiently and also feel like they could still be hacked. I am probably just paranoid.  Still makes me nervous, though."}
    \item \textit{"I try not to use the save password feature because of an exploit found not too long ago that allowed people to steal your passwords by using that feature. Lastpass is very easy to use and is a good way to store the passwords and it also lets you generate new passwords that are safe to use. It makes it so that you don't have to try and come up with your own hand crafted passwords."}
\end{itemize}

Perhaps surprisingly, our participants also expressed often concern about physical attacks, e.g., illegitimate persons accessing their devices or when sharing devices with other users. We noticed that many of our participants reported in their impressions of password managers that they only use them on non-shared and/or stationary devices, because otherwise another device user or thief could easily gain access to accounts using the saved passwords. Maybe interestingly, the received descriptions fit more to browser built-in password managers, which, in contrast to 3rd party managers, by default not protect the password storage with a \textit{master} password, which in the event of device sharing or device loss forms an additional protection against immediate password leakage and misuse. For some users, those concerns affect their password storage strategies, for instance, by restricting usage of password managers to less important websites. Prototypical answers that expressed those concerns were:

\begin{itemize}
    \item \textit{"I don't like using the browser to store them because I use the same browser across devices and wouldn't want someone to be able to get my passwords if my phone was stolen. I like using LastPass because it has a password protecting all of the other passwords. It feels safer than just saving them with the browser."}
    \item \textit{"I don't feel like it is the most secure. What if someone steals my laptop? But, I use it anyway, because the reward outweighs the risk. Most likely no one will steal my laptop. And I do not carry it around with me, it stays at home."}
    \item \textit{"My impression is fairly good using the browser. Unless my computer or phone is stolen and then hacked I'm pretty safe."}
    \item \textit{"It's probably not the safest thing in the world but I don't think my laptop will get stolen and no one else has access to my laptop soo..."}
    \item \textit{"I think for personal use, using a browser's password saving feature is fine. It's true that someone could steal the computer and get in, but I consider that to be an acceptable risk in the name of ease of use. [...]"}
    \item \textit{"I use it to automatically save some of my passwords on some sites so that I don't have to manually type it every time. It is useful, and I like it. My only concern is that it could provide unwanted access if my computer were stolen, but given how and where I use it this is fairly unlikely."}
    \item \textit{"I guess the browser built in password savers are password managers, but they are not secure. any number of programs can dump the passwords, but they are convenient for loggin~[sic] and it does save copying and pasteing~[sic]. I am the only one using this computer so I would have to scramble if it were every stolen or compromised"}
\end{itemize}

\subsubsection*{Misunderstood security benefits of password managers}

Lastly, we also noticed a very few cases that users attributed password managers security benefits that this software does not offer. For instance, one user noted that a password manager offers protection against password leakage through keyloggers (\textit{"Using my browser's password saving feature is a matter of convenient as it allows me to not have to always type in my password and it's a feature that I feel is fairly secure. Also, not having to type in a password, allows me to bypass keylogger and other risks that may be associated with frequent use of typing in password."}), which is only true in a limited set of scenarios (e.g., compromised keyboard firmware or USB keylogging devices on the keyboard USB port), but does not hold when the end-user device is compromised with malware. In the latter case, malware might steal the password database and log the user's master password for that database, if set.

\subsection{Threats to validity}

As with other human-subject and field studies, we cannot eliminate all threats to the validity of our study. We targeted Google Chrome users, which had in general~\cite{w3counter} the highest market share and also among our survey participants. Further, we recruited only experienced US workers on Amazon MTurk, which might not be representative for any population or other cultures (external validity), however, our demographics and password statistics show alignment with prior studies. Furthermore, we collected our data \textit{in the wild}, which yields a high ecological validity and avoids common problems of password lab studies~\cite{Komanduri:2011:PPM:1978942.1979321}, but on the downside does not give control over all variables (internal validity). We asked our participants to behave naturally and also tried to encourage this behavior through transparency, availability, and above average payment, however, like closest related work~\cite{197316,naturalhabitat} we cannot exclude that some participants behaved unusually.

%% file: sections/Appendix.tex
\section{Initial Survey Questions}
\label{sec:questionnaire}

\noindent\textit{\textbf{Q1:} For each of the following statements, how strongly do you agree or disagree?}\\
    a1: Consumer have lost all control over how personal information is collected and used by companies.\\
    a2: Most businesses handle the personal information they collect about consumers in a proper and confidential way. \\
    a3: Existing laws and organizational practices provide a reasonable level of protection for consumer privacy today.\\
(i)~Strongly disagree, (ii)~Somewhat disagree, (iii)~Somewhat agree, (iv)~Strongly agree \\

\noindent\textit{\textbf{Q2:} On how many different Internet sites do you have a user account that is secured with a password? (If you are not sure about the number please estimate the number)} (FreeText)\\

\noindent\textit{\textbf{Q3:} Has ever one of your passwords been leaked or been stolen?}\\
(i)~Yes, (ii)~No, (iii)~I am not aware of that, (iv)~I do not care \\

\noindent\textit{\textbf{Q4:} How strongly do you agree or disagree:?}

    b1. Passwords are useless, because hackers can steal my data either way. \\
    (i)~Strongly disagree, (ii)~Somewhat disagree, (iii)~Somewhat agree, (iv)~Strongly agree
    
    b2. I don't care about my passwords' strength, because I don't have anything to hide.\\ 
     (i)~Strongly disagree, (ii)~Somewhat disagree, (iii)~Somewhat agree, (iv)~Strongly agree\\
     
\noindent\textit{\textbf{Q5:} What characterizes in your opinion a strong/secure password?} (FreeText)\\

\noindent\textit{\textbf{Q6:} Please rate the strength of the following passwords?} \\	
    c1. thHisiSaSecUrePassWord \\
    c2. Pa\$sWordsk123 \\
    c3. AiWuutaiveep9j \\
    c4. !@\#\$\%\^\&*()\\ 
    c5. 12/07/2017 \\
    (i)~Very weak, (ii)~Weak, (iii)~Moderate strength, (iv)~Strong, (v)~Very strong \\

 \noindent\textit{\textbf{Q7:} I have never used a computer?} \\
 (i)~I have never, (ii)~I do \\
 
 \noindent\textit{\textbf{Q8:} How would you rate your ability to create strong passwords?} \\
 (i)~5(high ability), (ii)~4, (iii)~3, (iv)~2, (v)~1(low ability)~\\

 \noindent\textit{\textbf{Q9:} How do you proceed if you have to create a new password? (What is your strategy?)}  (FreeText)\\

  \noindent\textit{\textbf{Q10:} I try to create secure passwords.....} \\
  (i)~for all my accounts and websites, (ii)~for my email accounts, (iii)~for online shopping, (iv)~for online booking/reservation, (v)~for social networks, (vi)~No answer, (vii)~Other \\
  
  \noindent\textit{\textbf{Q11:} I make a point of changing my passwords on websites that are critical to my privacy every...... (choose the closest match)}	\\
  (i)~Day, (ii)~Week, (iii)~Two weeks, (iv)~Month, (v)~6 month, (vi)~Year, (vii)~Never, (viii)~Other \\

   \noindent\textit{\textbf{Q12:} Do you use the same password for different email accounts, websites, or devices?} \\
  (i)~Yes, (ii)~No \\
  
\noindent\textit{\textbf{Q13:} Do you use any of the following strategies for creating your password or part of your password, anywhere, at any time in the last year...} \\
 (i)~I used the name of celebrities as a password or as a part of a password, (ii)~I used the name of family members as a password or as a part of a password, (iii)~I used literature (book, poetry, etc.)~as a password or as a part of a password, (iv)~I used familiar numbers (street address, employee number, etc)~as a password or as a part of a password, (v)~I used random characters as a password, (vi)~I used a password manager to generate passwords, (vii)~No answer, (viii)~Other \\
 
 \noindent\textit{\textbf{Q14:} How do you remember all of your passwords?}	\\
  (i)~I write them down on paper (notebook, day planner, etc), (ii)~I try to remember them (human memory), (iii)~I use computer files (Word document, Excel sheet, text file, etc), (iv)~I use encrypted computer files (e.g. CryptoPad), (v)~I store my passwords on my mobile phone or PDA, (vi)~I use 3rd party password manager (save in extra program, e.g. LastPass, keepass, 1Password, etc.), (vii)~I use website cookies (Website checkbox: "Remember my password on this computer"), (viii)~I use the same password for more than one purpose, (ix)~I use browser built-in password manager (i.e saved in browser), (x)~I use a variation of a past password (eg. password1 and then password2 and then password3, etc.), (xi)~No answer, (xii)~Other \\

 \noindent\textit{\textbf{Q15:} Have you ever used a computer program to generate your passwords? } \\	
 (i)~Yes, (ii)~No \\
 
\noindent\textit{\textbf{Q16:} When creating a new password, which do you regard as most important: choosing a password that is easy to remember for future use (ease of remembering)~or the password's security? } \\
(i)~Always ease of remembering, (ii)~Mostly ease of remembering, (iii)~Mostly security, (iv)~Always security, (v)~Other \\

  \noindent\textit{\textbf{Q17:} When you create a new password, which of the following factors do you consider? The password ....} \\
   (i)~does not contain dictionary words, (ii)~is in a foreign (non-English)~language, (iii)~is not related to the site (i.e., the name of the site), (iv)~includes numbers, (v)~includes special characters (e.g. "\&" or "!"), (vi)~is at least eight (8)~characters long, (vii)~None of the above: I didn't think about it, (viii)~No answer, (ix)~Other \\
   
\noindent\textit{\textbf{Q18:} My home planet is Earth?}	\\ 
 (i)~Yes, (ii)~No \\
 
 \noindent\textit{\textbf{Q19:} Do you use the "save password" feature of your browser? } \\
 (i)~Yes, (ii)~No \\
 
  \noindent\textit{\textbf{Q20:} Do you use any kind of extra password manager program (for instance, LastPass, 1Password, Keepass, Dashlane, etc.)?} \\
 (i)~Yes, (ii)~No \\
 
\noindent\textit{\textbf{Q21:} Which password manager(s)~do you use? (You can write one name per line)}  (FreeText)\\

 \noindent\textit{\textbf{Q22:} Please give us a short description of your impression of using your browser's password saving feature and/or of using extra password managers}  (FreeText)\\

 \noindent\textit{\textbf{Q23:} How many passwords do you keep in your password manager(s)~and browser's saved passwords? (if you don't know the exact number, please estimate the number)}  (FreeText)\\

 \noindent\textit{\textbf{Q24:} I am } (i)~Female, (ii)~Male, (iii)~Other, (iv)~No answer\\

 \noindent\textit{\textbf{Q25:} My age group is	 } (i)~under 18 years, (ii)~18 to 30, (iii)~31 to 40, (iv)~41 to 50, (v)~51 to 60, (vi)~61 to 70, (vii)~71 or older, (viii)~Other \\
 
 \noindent\textit{\textbf{Q26:} My native language is} (FreeText)\\
   
 \noindent\textit{\textbf{Q27:} My primary web browser is} (i)~Chrome, (ii)~Firefox, (iii)~Internet explorer/ Edge, (iv)~Safari, (v)~Opera, (vi)~Other\\
  
  \noindent\textit{\textbf{Q28:} For browsing websites, I use} (i)~Almost exclusively my smartphone/tablet, (ii)~Mostly my smartphone / tablet, (iii)~Almost exclusively my desktop / laptop computer, (iv)~Mostly my desktop / laptop computer \\
  
 \noindent\textit{\textbf{Q29:} What is the highest degree or level of education you have completed?} \\
 (i)~Less than high school, (ii)~High school graduate (includes equivalency), (iii)~Some collage/no degree, (iv)~Associate's degree, (v)~Bachelor's degree, (vi)~Ph.D, (vii)~Graduate or professional degree, (viii)~Other \\
 
 \noindent\textit{\textbf{Q30:} Are you majoring in or do you have a degree or job in computer science, computer engineering, information technology, or a related field?} \\
  (i)~Yes, (ii)~No \\
  
 \noindent\textit{\textbf{Q31:} What is your ethnicity?} \\
 (i)~White/Caucasian, (ii)~Black/African American, (iii)~Asian, (iv)~Hispanic/Latino, (v)~Middle Eastern, (vi)~Native American/Alaska native, (vii)~Native Hawaiian/Pacific Islander, (viii)~Multiracial, (ix)~Other \\

\section{Exit Survey Questions}
\label{sec:exitsurvey} 

\noindent\textit{\textbf{ES1:} I do not use any kind of 3rd party password manager, such as 1Password, LastPass, etc,. because?} \\
 (i)~I do not trust the password manager software or vendor, (ii)~of the lack of support for my devices, (iii)~of the lack of synchronization between different devices, (iv)~I would have to spent money on it, (v)~they are not simple to set up and/or not easy to use, (vi)~I can manage my passwords myself and a password manager would not provide any additional benefits, (vii)~Chrome's password saving feature suffices for me, (viii)~there are too many available managers and I am not sure which one would be right for me, (ix)~I never really thought about using a 3rd party password manager or was never interested in them, (x)~Other \\
 
 \noindent\textit{\textbf{ES2:} Have you ever used any kind of 3rd party password manager in the past, and then stopped using it?}	 \\
 (i)~Yes, (ii)~No \\
 
 \noindent\textit{\textbf{ES3:} Please mention which 3rd party password manager have you used in the past.} (FreeText)\\
 
 \noindent\textit{\textbf{ES4:} Please shortly explain the reasons why you stopped using the 3rd party password manager.} (FreeText)\\